\documentclass[10pt,a4paper,nofootinbib]{article}
\usepackage{graphicx}
\pdfoutput=1
\usepackage{jheppub}
\usepackage{parskip}
\usepackage{amsmath,amsfonts,amssymb,bbm, mathtools, mathrsfs}
\usepackage{hyperref, csquotes}
\usepackage{graphicx, color, svg,subcaption}
\usepackage{tikz,pgf,tikz-cd,float}
\usepackage{physics,tensor}
\usepackage{rotating}
\usepackage{todonotes}
\usepackage{braket}
\usepackage{amsmath}
\usepackage{bm}
\usepackage{amssymb}
\usepackage{xcolor}
\usepackage{multicol}
\usepackage{multirow}
\usepackage[normalem]{ulem}
\usepackage{comment}
\usepackage{lmodern}

\newcommand{\e}{\textrm{e}}
\renewcommand{\d}{\text{d}}

\title{Gauge versus (hidden) physical symmetries of FLRW cosmologies}
\author[a]{Andrea Calcinari,}
\author[a,b]{Adri\`a Delhom,}
\author[c,d]{Federico Greco,}
\author[a]{Daniele Oriti,}
\author[a]{N\'estor Rivero}

\affiliation[a]{Departamento de F\'isica Te\'orica, Facultad de Ciencias F\'isicas, Universidad Complutense de Madrid, Plaza de las Ciencias 1, 28040 Madrid, Spain}
\affiliation[b]{Laboratoire Kastler Brossel, Sorbonne Université, CNRS, ENS-Université PSL, Collège de France, 4 Place Jussieu, 75005 Paris, France}
\affiliation[c]{Dipartimento di Fisica e Astronomia “G. Galilei”, Universit\`a degli Studi di Padova, via Marzolo 8, I-35131 Padova, Italy}
\affiliation[d]{INFN, Sezione di Padova, via Marzolo 8, I-35131 Padova, Italy}

\emailAdd{andrcalc@ucm.es}
\emailAdd{adria.delhom@lkb.upmc.fr}
\emailAdd{federico.greco.1@phd.unipd.it}
\emailAdd{doriti@ucm.es}
\emailAdd{nestoriv@ucm.es}

\date{\today}

\abstract{In generally covariant theories, evolution in coordinate time is a gauge transformation, so that a symmetry made manifest in a gauge-fixed description need not be a symmetry of the physical dynamics. At the same time, deparametrisation removes gauge redundancies but may hide physical symmetries, in particular those that depend on the chosen physical clock. We investigate the relation between gauge and hidden physical symmetries in flat FLRW geometry coupled to an arbitrary number $n$ of free massless scalar fields. We show that conformal Killing vectors of the minisuperspace metric generate conserved charges that are Dirac observables and hence gauge-invariant. Their Poisson algebra realises the maximal conformal algebra $\mathfrak{conf}(n,1)\simeq\mathfrak{so}(n+1,2)$, extending previous single-field results to arbitrary $n$. We then revisit the Eisenhart--Duval lift in a family of gauges and show that the resulting symmetry algebra is gauge dependent. Only in the harmonic gauge does the algebra enlarge to the Schr\"odinger algebra, which is thus not a physical symmetry. Finally, we show that deparametrisation maps the lifted charges to gauge-invariant Dirac observables, which always realise a subalgebra of $\mathfrak{conf}(n,1)$, recovering it in full in the harmonic gauge. Our results provide a systematic framework for disentangling gauge from physical symmetries in minisuperspace models, recovering charges to which reduced phase-space descriptions are structurally blind, and extending naturally to models with potentials.
}

\begin{document}

\maketitle

\section{Introduction}\label{sec:intro}

General relativity is invariant under spacetime diffeomorphisms, and its canonical formulation yields a time-reparametrisation invariant constrained system. As a result, the Hamiltonian is a combination of first-class constraints, so that evolution in coordinate time is a gauge transformation rather than physical change~\cite{Dirac:1958sc,Arnowitt:1962hi}. The true dynamical content of the theory stems from correlations between physical quantities, so that the dynamics is relational~\cite{Rovelli:2001bz,Dittrich:2004cb,CarloBook,ThiemannBook}. 

While the canonical dynamics is encoded on the constrained phase space, the configuration space inherits a natural geometric structure through the DeWitt supermetric appearing in the Hamiltonian constraint. In the symmetry-reduced models considered here, the classical solutions correspond to null geodesics of this supermetric, making it natural to ask which of its geometric symmetries correspond to symmetries of the physical, relational dynamics. Continuous symmetries can be exploited to organise and classify the space of classical solutions, and
generate the conserved charges that label them. Given that coordinate time is gauge, not every symmetry one may find in a time-dependent, gauge-fixed description of the system's dynamics need be physical in this sense---distinguishing genuine symmetries of the physical dynamics from symmetries of a particular gauge choice is therefore essential to any such classification.

Although the relational formalism applies in principle to the full superspace, an explicit closed-form symmetry classification of the kind we pursue here is obstructed in that setting. Superspace is a stratified space rather than a smooth manifold, singular precisely at the three-geometries admitting continuous isometries~\cite{Giulini1995,Giulini:2009np}. Moreover, the DeWitt supermetric is ultralocal on configuration space, but the potential-like term induced from the spatial curvature contribution is not, as it depends on spatial derivatives of the spatial metric up to second order. As a result, the equations governing its dynamical symmetries couple
neighbouring points of the spatial geometry through this term, rather than reducing to the finite-dimensional system of partial differential equations one
faces in minisuperspace. Symmetry-reduced models evade both obstructions by construction, which is why an explicit symmetry classification has so far only been carried out in such settings~\cite{sl2r1,EteraDaniele,Chiba:2024auh,Chiba:2024iia,Sano:2025xit,Ribisi:2024tmk,Geiller:2020xze,Geiller:2022baq}. Beyond these mathematical simplifications, such models are also interesting in their own right, and their symmetry structure sheds light on the solvability and algebraic structure of the reduced gravitational phase space, and is directly relevant to its quantisation.

The dynamical symmetries of minisuperspace models have accordingly received considerable attention. On the one hand, Hamilton--Jacobi methods have been applied to construct constants of motion and integrate classical trajectories in time-reparametrisation invariant cosmologies~\cite{vanHolten:2013yca,Bogers:2015fwa}. On the other hand, field-space geometric methods have been used to uncover hidden symmetry algebras. In cosmic time, flat FLRW cosmology coupled to free massless scalar fields has been shown to possess a hidden $\mathfrak{sl}(2,\mathbb{R})$ conformal symmetry~\cite{sl2r1, sl2r2,sl2r3}, later extended, via the Eisenhart--Duval lift~\cite{ED,ED2,ED3}, to a full Schr\"odinger symmetry for a single scalar field~\cite{EteraDaniele}. A cosmological extension of the lift itself, incorporating the scale factor and the energy-momentum tensor, has also been developed~\cite{Cariglia:2018mos}. The same lift technique has been used to classify conformal Killing vectors of the single-field minisuperspace for a general potential~\cite{Chiba:2024auh}, and to identify the exponential potentials responsible for the hidden symmetry of power-law inflation~\cite{Chiba:2024iia}. Two aspects of this body of work remain open. First, these results are confined to a single scalar field and to flat FLRW spacetimes. The extension to several matter fields, or to nonvanishing spatial curvature, has so far not been worked out~\cite{Chiba:2026chq}; several massless scalars have been treated
in the different setting of spherically symmetric static
minisuperspaces~\cite{Sano:2025xit}. Second, and more important, in \textit{all} of these constructions the symmetry algebra is obtained in fixed gauges, and it is not clear which part of the resulting structure is a property of the physical system and which part is an artefact of the gauge choice.

In this paper we set up a framework to study the symmetries of minisuperspace models within their relational formulation, designed precisely to make this distinction manifest. We work out explicitly how symmetries obtained in a gauge-dependent description relate to the gauge-invariant, relational content of the theory, and identify where the two conflate. We apply this framework in full detail to flat FLRW cosmology coupled to $n$ free massless scalar fields, extending existing single-field results to arbitrary $n$ and using the model to exemplify, concretely, how gauge and physical symmetries can be told apart.

We proceed in two steps. First, working directly on minisuperspace (using the geometrisation of cosmological dynamics as the motion of a point particle on minisuperspace) and without reference to any time coordinate, we exploit the fact that, in the absence of a potential, classical trajectories are null geodesics of the field-space supermetric. Because conformal transformations preserve null cones, the conformal Killing vectors generate the associated conserved charges, which we construct explicitly as weak Dirac observables. We observe that, upon deparametrisation, the clock momentum acts as an effective mass, trading the null cone for a mass shell, so that a reduced description makes manifest only isometries while the relationally (clock) time-dependent charges become hidden. This motivates our second step: we apply the Eisenhart--Duval lift, which geometrises explicit time dependence using higher-dimensional null cones \cite{ED,ED2,ED3}, performing the construction for an arbitrary lapse. We solve the resulting conformal Killing equations on a two-parameter family of gauges rather than in a single gauge like in previous work, so as to track explicitly how the manifest symmetry algebra depends on the choice of time gauge, and how it relates, by deparametrisation, to the gauge-invariant charges found in the first step.

This approach yields a complete, gauge-independent classification of the dynamical symmetries of the model that are generated by charges linear in the momenta, and directly addresses two open issues identified above: inclusion of $n$ matter fields and disentangling of gauge and physical symmetries. For an FLRW geometry with $n$ scalar fields (thus also for the Bianchi I universe, whose two anisotropies enter the reduced action precisely as free massless scalar fields \cite{BojoBook}), the conformal Killing vectors of the supermetric saturate the maximal conformal algebra $\mathfrak{conf}(n,1)$. Working with the Eisenhart--Duval lift in a family of gauges, we show that this algebra becomes hidden. The lifted charges close only into $\mathfrak{sl}(2,\mathbb{R})\oplus\mathfrak{iso}(n)$, which enlarges to a centrally extended Schr\"odinger algebra $\widehat{\mathfrak{sh}}(n,1)$ in a single (harmonic) gauge, where the gauge-fixed minisuperspace metric becomes flat. Upon deparametrisation, the picture becomes gauge-invariant: in every gauge the resulting charges are contained in the conformal algebra obtained in the first step, and they exhaust it precisely in the harmonic gauge---recovering then even the relationally time-dependent charges lost in clock-reduced descriptions. The Schr\"odinger enhancement is therefore recovered as a property of one specific gauge, rather than as a physical symmetry, resolving the gauge-versus-physical ambiguity left open in previous constructions. We also observe that the single-field case is special in this respect: for $n=1$ the gauge-fixed metric is flat in \emph{every} gauge, so the enhancement persists as an accident of two dimensions,
rather than a property of the gauge in which it was found \cite{EteraDaniele}.

The paper is organised as follows. Section~\ref{sec:Constrained} reviews time-reparametrisation-invariant systems, relational dynamics via deparametrisation, and the construction of relational Dirac observables, specialising later to a configuration space equipped with a field-space metric. Section~\ref{sec:FLRW} constructs the relational dynamics and the full conformal algebra of weak Dirac observables for FLRW cosmology with $n$ scalar fields, directly and without reference to any time gauge. Section~\ref{sec:Lift} applies the Eisenhart--Duval lift for a family of lapses, tracks explicitly how the manifest symmetry algebra depends on the gauge, identifies the distinguished harmonic gauge in which it enlarges to a Schr\"odinger algebra, and recovers the gauge-invariant conformal charges of Section~\ref{sec:FLRW} by deparametrisation. Section~\ref{sec:outlook} collects our conclusions and outlines directions for future work. Appendix~\ref{app:ED} reviews the Eisenhart--Duval lift, keeping the lapse arbitrary throughout, and Appendix~\ref{app:reconstruction} shows how to reconstruct the relational trajectories of Section~\ref{sec:FLRW} algebraically from the conserved charges.

\section{Time-reparametrisation invariance and relational observables}
\label{sec:Constrained}

In this paper, we restrict our attention to symmetry-reduced homogeneous GR coupled to massless scalar fields. Spatial dependence then drops out and the system reduces to a finite-dimensional mechanical model that is invariant under time reparametrisations \cite{BojoBook}. 

The dynamics of a generic time-reparametrisation invariant mechanical system can be described by an action of the form
\begin{equation}
    \label{eq:General_action}
    S=\int_{\mathbb{R}} \d \lambda \,{\mathcal{L}}(\phi^a,\dot \phi^a)\,,
\end{equation}
where the Lagrangian satisfies ${\mathcal{L}}(\phi^a,\alpha\dot \phi^a)=\alpha{\mathcal{L}}(\phi^a,\dot{\phi}^a)$. Here $\phi^1,\dots,\phi^d$ are the $d$ degrees of freedom of the system, $\dot{\phi}^a\coloneqq \d \phi^a/\d\lambda$ are the associated velocities, and $\lambda$ is a generic time parameter.

Homogeneity in the velocities guarantees that the action \eqref{eq:General_action} is invariant  under a time reparametrisation $\lambda\rightarrow f(\lambda)$. On the other hand, it also implies that the Hessian with respect to the velocities has at least one null direction (leading to a degenerate Legendre transform), and that the canonical Hamiltonian identically vanishes $H_c=0$. If the rank of the Hessian is $d-1$, there is a single primary constraint $h(\phi^a, p_a)\approx0$\footnote{Here and throughout, $\approx$ denotes the weak equality of \cite{Diracbook}.} known as Hamiltonian constraint, where $p_a$ are the canonical momenta conjugated to $\phi^a$. In this case, the total Dirac Hamiltonian $H=H_c+u(\lambda)h$ is \cite{Diracbook,Gaugebook}
\begin{align}\label{eq:form Hamiltonian}
     H(\phi^a,p_a)=u(\lambda) \,h(\phi^a,p_a) \,,    
\end{align} 
where $u(\lambda)$ is a multiplier\footnote{In general relativity, this Lagrange multiplier is called \textit{lapse} function and is usually denoted $N(\lambda)$.} which enforces the primary constraint $h\approx0$. This defines a ($2d-1$)-dimensional surface ${\mathcal{C}_H}$ in phase space $\mathcal{P}$. If we introduce the time parameter
\begin{equation}
\tau(\lambda)=\int_0^\lambda \d\xi\,u(\xi)\,,
    \label{eq:tau_lambda}
\end{equation} 
evolution is described via the phase space Hamiltonian flow $\alpha_h^\tau:\mathcal P\rightarrow\mathcal P$ generated by $h$, whose action is encoded in the Poisson structure. If $x_0\in\mathcal P$ is an initial point, then $x(\tau)=\alpha_h^\tau(x_0)$ is the corresponding phase-space trajectory. In turn, for any phase-space function $f\in\mathcal{F}(\mathcal{P})$, where $\mathcal{F}(\mathcal{P})$ denotes the algebra of smooth real-valued functions on $\mathcal{P}$, the composition $f(x(\tau))$ satisfies
\begin{equation}
    \label{eq:Hamiltonian_flow_tau}
    \frac{\d f(x(\tau))}{\d\tau}=\left.\{f,h\}\right|_{x(\tau)} \,.
\end{equation}
This time parametrisation allows for a simple expression of the action of the Hamiltonian flow on phase-space functions, which we denote by $A_h^\tau:\mathcal{F}(\mathcal{P})\to\mathcal{F}(\mathcal{P})$, and can be written as \cite{DittrichDO,GieselDO}
\begin{equation}
    \label{eq:Hamiltonian_flow_alpha}
    A_h^\tau(f)
    =
    \sum_{n=0}^{\infty}
    \frac{\tau^n}{n!}
    \{f,h\}_n \,,
\end{equation}
where we have defined $\{f,h\}_0=f$ and $\{f,h\}_{n+1}=\{\{f,h\}_n,h\}$. Evaluating this expression at $x_0\in\mathcal{P}$ then provides the solution to Eq.~\eqref{eq:Hamiltonian_flow_tau}. In terms of the general time parameter $\lambda$ used above, the same Hamiltonian evolution described by Eq.~\eqref{eq:Hamiltonian_flow_tau} reads
\begin{equation}
    \label{eq:Hamiltonian_flow}
    \frac{\d f(x(\lambda))}{\d\lambda}=\left.\{f,H\}\right|_{x(\lambda)}=\left.u(\lambda)\{f,h\}\right|_{x(\lambda)}\,.
\end{equation}
The multiplier therefore does not change the geometric orbit generated by the Hamiltonian constraint $h$, but only its parametrisation. Importantly, the Hamiltonian vector field $\{\,\cdot\,,h\}$ generating the orbits is tangent to the constraint surface $\mathcal{C}_H$ ($\{\,h\,,h\}=0$), so that any orbit intersecting $\mathcal{C}_H$ is entirely contained in it. In other words: if an initial condition satisfies the (first-class) constraint $h\approx 0$, the corresponding trajectory remains entirely within $\mathcal{C}_H$, which is therefore left invariant under the Hamiltonian flow \cite{Wipf:1993xg}.

Since the action \eqref{eq:General_action} is invariant under time reparametrisations, the evolution generated by the Hamiltonian constraint is not physical, but rather corresponds to a gauge transformation along $\mathcal C_H$ (where different choices of time parameter simply provide different parametrisations of the same orbit). Importantly, because any physical observable $Q$ must be gauge-invariant, it must be constant along the gauge flow. More precisely, it must be a real-valued phase-space function
satisfying
\begin{equation}
    \label{eq:Dirac_observable}
    \{Q,h\}\approx 0\,,
\end{equation}
where $\approx$ holds on the constraint surface $\mathcal C_H$ but not necessarily on the whole phase space $\mathcal{P}$. If \eqref{eq:Dirac_observable} holds, $Q$ is a \textit{weak Dirac observable}. If instead one has 
\begin{equation}\label{eq:strong}
    \{Q,h\}=0 
\end{equation}
on the full phase space, $Q$ is a \textit{strong Dirac observable}. Dirac observables are constant along the gauge flow \cite{Kuchar:1991qf,Isham:1992ms}, but can be used to describe dynamics in a gauge-invariant (relational) way \cite{DittrichDO,GieselDO,Tambornino}: evolution must be defined with respect to an internal dynamical degree of freedom of the system rather than an arbitrary gauge parameter. This leads to \textit{relational observables}, where dynamics are parametrised by an internal degree of freedom used as relational clock. In general, such a clock can be chosen to be any \textit{physical}\footnote{While this abstract definition allows for highly nontrivial phase-space combinations (e.g., mixtures of variables and momenta), practical implementations usually specialise $T$ to be a configuration coordinate, as detailed below.} phase-space function $T$ (serving as a \textit{partial} observable \cite{RovelliPO,DittrichDO,Dittrich:2005kc}) that provides a well-defined parametrisation of the gauge orbit, which requires $\{T, h \} \not\approx 0$. 

Given a phase-space function $f$ and a chosen relational clock $T$, there is a canonical procedure to build a relational Dirac observable $Q_{f,T}(t)$ relying on the strategy of \textit{evolving constants of motion} \cite{Rovelli_Amodel,Rovelli_Anhypothesis,RovelliOBS,RovelliPO}. First, one solves the equation $A_{h}^\tau (T)= t$ for the flow parameter to obtain $\tau_T(t)$. Then, one can define the relational Dirac observable as \cite{Trinity,RelativisticTrinity}
\begin{equation}
    \label{eq:Relational_observables}
    Q_{f,T}(t) := A_{h}^\tau (f) \big \vert_{\tau=\tau_T(t)} \approx \sum_{n=0}^{\infty} \frac{(t-T)^n}{n !} \left\{ f, \frac{h}{\{T,h\}}  \right\}_n\,,
\end{equation}
where we used $\{T, \frac{h}{\{T,h\}}\} \approx 1$. Provided that $T$ is a good relational clock (i.e., $\{T,h\} \not\approx0$), one can check that $Q_{f,T}(t)$ is a  gauge-invariant (weak) Dirac observable, whose physical interpretation is \textit{the value of $f$ when the relational clock $T$ reads the value $t$}. Such an evolving constant of motion is called \textit{complete} observable, to distinguish it from the {partial} observables such as $f$ and $T$ \cite{RovelliPO,DittrichDO,Dittrich:2005kc}.

While this framework constructs relational observables directly on the full phase space for a generic physical clock $T$, the content of such gauge-invariant dynamics can be completely isolated through the procedure of \textit{deparametrisation} \cite{Tambornino}. Take the clock $T$ to be one of the internal configuration degrees of freedom, $T \coloneqq \phi^c$, i.e., a genuine dynamical variable with conjugate momentum $p_T \coloneqq p_c$. Exploiting the fact that $\{T,h\} \not\approx 0$, one can locally solve the Hamiltonian constraint $h \approx 0$ for this specific clock momentum $p_T$ in terms of a function $H_r$ which depends only on $(\phi^A,p_A,T)$, with $A$ running over all configuration indices $a$ except the clock index. Importantly, $H_r$ generates evolution with respect to this relational clock, thus serving as the relational (or physical) Hamiltonian within the reduced phase space description. Deparametrisation is thus achieved by implementing the Hamiltonian constraint $h\approx 0$ as
\begin{equation}
    \label{eq:deparam_constraint}
    p_T \approx -H_r(\phi^A,p_A,T) \,.
\end{equation}
This linear form arises naturally for ideal non-relativistic clocks \cite{Trinity}. For relativistic systems where the clock momentum appears quadratically, on the other hand, this form is achieved by taking a square root of the quadratic constraint and restricting the evaluation to a single frequency sector, thus a direction in the internal time \cite{RelativisticTrinity}.

This general setup allows us to define a relational expression $f_r$ for any phase-space function $f$ by evaluating it on the constraint surface:
\begin{equation}
\label{eq:DeparametricePhaseSpaceFunction}
    f_r(\phi^A,p_A,T)\coloneqq f\left(\phi^A,p_A,T,-H_r(\phi^A,p_A,T)\right)\,.
\end{equation}
Importantly, through this map, any weak Dirac observable $Q$ leads to a conserved quantity along deparametrised (relational) trajectories $Q_r(\phi^A,p_A,T)$. Since $\{Q,h\}\approx0$ and $T$ is a valid clock ($\{T,h\} \not\approx 0$), its total rate of change with respect to the relational clock vanishes weakly:
\begin{equation}
    \frac{\d Q}{\d T} = \frac{\{Q,h\}}{\{T,h\}} \approx 0\,.
\end{equation}
Equivalently, its conservation along the trajectories that solve the relational Hamiltonian dynamics is given by
\begin{equation}\label{eq:Qr_Conserved_wrt_Any}
    \frac{\d Q_r}{\d T}=\frac{\partial Q_r}{\partial T}+\{Q_r,H_r\}_\text{red}\approx0\,,
\end{equation}
where the Poisson brackets $\{\cdot,\cdot\}_\text{red}$ are associated with the symplectic structure of the reduced phase space spanned by $(\phi^A,p_A)$.

A notable strength of Eq.~\eqref{eq:Qr_Conserved_wrt_Any} is that it remains structurally valid regardless of which internal degree of freedom is chosen as the relational clock $T$ (with corresponding relational Hamiltonian $H_r \approx -p_T$). Consequently, once a set of Dirac observables $Q$ is identified on the full unreduced phase space $\mathcal{P}$, their reduced counterparts $Q_r$ are guaranteed to operate as physical, relationally conserved charges. This remark is crucial for our analysis because it eliminates the need to deparametrise explicitly for each specific degree of freedom prior to exploring the system's symmetries, thereby avoiding the clock-dependent calculations inherent to the reduced phase space description.

Interestingly, the relational observable \eqref{eq:Relational_observables} simplifies greatly if we restrict our attention to a relational Hamiltonian $H_r = H_r(\phi^A, p_A)$ and to functions $f_r = f_r(\phi^A, p_A)$ that carry no explicit clock dependence. This is a standard simplifying assumption in the literature of relational dynamics \cite{Trinity,RelativisticTrinity}, where the total phase space is assumed to decouple cleanly into a product of a clock sector and a non-clock sector. Physically, this implies that the clock and the rest of the system do not dynamically interact, and since the relational Hamiltonian depends exclusively on the non-clock variables, all Poisson brackets can be evaluated directly on the reduced phase space.\footnote{Because $f_r$ and $H_r$ are independent of the clock, the full phase space Poisson brackets coincide with the symplectic structure $\{\cdot,\cdot\}_\text{red}$ of the reduced phase space, which is spanned by non-clock degrees of freedom.} Then, \eqref{eq:Relational_observables} reduces to the simple form \cite{Trinity,RelativisticTrinity}
\begin{equation}
\label{eq:Relational_observables_simplified}
    Q_{f_r,T}(t) \approx \sum_{n=0}^{\infty} \frac{(t-T)^n}{n !} \left\{ f_r, H_r\right\}_n\,,
\end{equation}
which describes the evolving constant of motion associated with a function $f_r$ of the reduced phase space, tracking its relational trajectory with respect to the partial observable $T$. As we will see later, the simple expression \eqref{eq:Relational_observables_simplified} provides a convenient tool to calculate relational dynamics in the homogeneous cosmologies of interest to us, allowing for instance to track the evolution of the volume of an FLRW universe with respect to a free massless scalar field used as relational clock.

To conclude the section, we outline the operational link between these gauge-invariant structures and the original coordinate time parameter. A convenient way to ensure that the time parameter tracks evolution at the exact same rate as the chosen internal degree of freedom is to fix the multiplier $u(\lambda)$ (see \eqref{eq:form Hamiltonian} and \eqref{eq:Hamiltonian_flow}) on the constraint surface $\mathcal{C}_H$ to the clock-adapted function 
\begin{equation}
\label{eq:u_multipl}
    u_T\big|_{\mathcal{C}_H}=\left.\frac{1}{\{T,h\}}\right|_{\mathcal{C}_H}\,,
\end{equation}
so that the gauge parameter ticks synchronously with the physical clock. However, it is important to note that this corresponds to a gauge fixing on the original unreduced phase space, hence it is not the same as properly deparametrising, as explained above. While $u$ fixes a parametrisation of the gauge orbit, a relational clock is a physical degree of freedom used to label points along that orbit. Thus, a choice of relational clock identifies a preferred adapted multiplier (or lapse function), but a generic lapse need not correspond to a globally well-defined internal clock. Crucially, the resulting relational observables themselves remain gauge-invariant and thus entirely independent of the original choice of multiplier, which is thus just a matter of convenience.

\subsection{Constrained mechanics with a field space metric}
\label{subsec:Supermetric}

So far we have worked with the homogeneous action \eqref{eq:General_action}, in which reparametrisation invariance is encoded in the degree-one homogeneity of the Lagrangian. Here we trade this for the equivalent \emph{lapse} formulation, in which the invariance is realised through an
auxiliary field $N$. A physically relevant class of models conforming to the general framework described above arises when the kinetic structure of the configuration space is governed by a field-space metric, often referred to as kinetic metric or \textit{supermetric} (also known as DeWitt metric in  gravitational contexts).  

Consider a mechanical system described by $d$ dynamical variables $\phi^a$ where configuration space is equipped with a metric $g_{ab}(\phi)$ of arbitrary signature, and subject to a potential $U(\phi)$. Borrowing the nomenclature of general relativity, we refer to the multiplier $u(\lambda)$ introduced previously as \textit{lapse function} $N(\lambda) $. The dynamics of such an inherently reparametrisation-invariant theory is governed by the action\footnote{This formulation is equivalent to that of the previous section, since eliminating $N$ via its equation of motion returns a Lagrangian that is homogeneous of degree one in the velocities. What changes is the canonical structure: the canonical Hamiltonian is $H_c=Nh$ and the primary constraint is instead $p_N\approx0$, with $h\approx0$ arising as a secondary constraint from its conservation. Both descriptions share the same constraint surface and gauge orbits.}
\begin{equation}
    \label{eq:Supermetric_action}
    S = \int_{\mathbb{R}} \d \lambda \, \left[ \frac{1}{2 N(\lambda)} g_{ab}(\phi) \dot{\phi}^a \dot{\phi}^b - N(\lambda) U(\phi) \right] \,.
\end{equation}
This specific action structure can be constructed by parametrising a standard mechanical system, but it naturally also emerges in frameworks that intrinsically lack a preferred background time. Most notably, this serves as the foundational formulation for general relativity and the symmetry-reduced cosmological models we will investigate in the following sections.

Within this class of systems, we restrict our attention to models where the potential vanishes identically, $U(\phi) = 0$. This restriction is physically well-justified for applications in homogeneous cosmology with no curvature potential and where matter is modelled by free, massless fields. Variation with respect to the dynamical fields $\phi^a$ yields the free geodesic equation
\begin{equation}
    \label{eq:EOM_phi_null}
    \frac{\d}{\d \lambda} \left( \frac{\dot{\phi}^a}{N} \right) + \frac{1}{N} \Gamma^a_{bc} \, \dot{\phi}^b \dot{\phi}^c = 0 \,,
\end{equation}
where $\Gamma^a_{bc}$ are the Christoffel symbols associated with the field space metric $g_{ab}$. On the other hand, because $N$ plays the role of a Lagrange multiplier, varying with respect to the lapse yields the constraint:
\begin{equation}
    \label{eq:EOM_N_null}
    g_{ab}(\phi) \dot{\phi}^a \dot{\phi}^b = 0 \,.
\end{equation}
This indicates that the velocity vector $\dot{\phi}^a$ must be null everywhere along the trajectory (however note that nontrivial dynamics requires an indefinite supermetric since for a Riemannian supermetric \eqref{eq:EOM_N_null} would force $\dot\phi^a=0$). If we map this to the reparametrisation-invariant time parameter $\tau$ defined via Eq.~\eqref{eq:tau_lambda} (such that $\d\tau = N \d\lambda$), Eq.~\eqref{eq:EOM_phi_null} reduces directly to the standard, affinely parametrised geodesic equation. The classical paths of this system are therefore precisely the null geodesics of the field space geometry.

The canonical analysis brings out the structural simplicity of this free formulation while tying it directly back to the gauge machinery introduced earlier. The canonical momenta conjugate to the coordinates $\phi^a$ are
\begin{equation}
    \label{eq:momenta_supermetric}
    p_a = \frac{\partial \mathcal{L}}{\partial \dot{\phi}^a} = \frac{1}{N} g_{ab} \dot{\phi}^b \,,
\end{equation}
while the momentum conjugate to the lapse vanishes, forming the primary constraint of the theory, $p_N \approx 0$. The canonical Hamiltonian then takes the form
\begin{equation}
    \label{eq:Canonical_Hamiltonian_Supermetric}
H = N(\lambda) \left[ \frac{1}{2} g^{ab}(\phi) p_a p_b \right] = N(\lambda) \, h (\phi^a, p_a) \,.\end{equation}
Demanding that $p_N$ be preserved along the evolution,
$\dot p_N = \{p_N, H\} 
= -h \approx 0$, yields the secondary Hamiltonian constraint defining the surface $\mathcal{C}_H$ as
\begin{equation}
    \label{eq:Hamiltonian_Constraint_Supermetric}
    h = \frac{1}{2} g^{ab}(\phi) p_a p_b \approx 0 \,.
\end{equation}
Because the inverse supermetric $g^{ab}$ defines the kinetic structure of the Hamiltonian constraint, the gauge flow equations on phase space project down to configuration space paths that are precisely the null geodesics of the field space. Consequently, the relational evolution of any observable via Eq.~\eqref{eq:Relational_observables} is fundamentally determined by this underlying field-space geometry.

To extract physical (relational) evolution, we apply the deparametrisation procedure introduced earlier. Recalling our field space split, we designate one of the configuration variables as the internal clock, $T \coloneqq \phi^c$, with conjugate momentum $p_c$, while the remaining physical degrees of freedom are denoted by the indices $A, B$. From this point forward, we focus on systems with \emph{diagonal} minisuperspace metric, meaning all cross-terms between the clock and the remaining physical variables vanish identically ($g_{cA} = 0$). Given this diagonal structure, which naturally arises in the homogeneous cosmological minisuperspaces of interest to us, the Hamiltonian constraint \eqref{eq:Hamiltonian_Constraint_Supermetric} takes the form
\begin{equation}
    \label{eq:diagonal_constraint}
    h = \frac{1}{2} \left( g^{cc} p_c^2 + g^{AB} p_A p_B \right) \approx 0 \,.
\end{equation}
To completely deparametrise the theory and cast the constraint into the linear form $p_c + H_r \approx 0$ of Eq.~\eqref{eq:deparam_constraint}, we must algebraically isolate the clock momentum $p_c$. To do so, we define a reduced inverse metric tensor $\gamma^{AB}$ on the non-clock configuration space as
\begin{equation}
    \label{eq:gamma_definition}
    \gamma^{AB} \coloneqq -\frac{g^{AB}}{g^{cc}} = -g_{cc}g^{AB} \,,
\end{equation}
where we used $g_{cA}=0$, so that we can rearrange Eq.~\eqref{eq:diagonal_constraint} explicitly as
\begin{equation}\label{eq:massivegeo}
    p_c^2 - \gamma^{AB} p_A p_B \approx 0 \,.
\end{equation}
Solving this for the clock momentum yields $p_c \approx \mp \sqrt{\gamma^{AB} p_A p_B}$, and using \eqref{eq:deparam_constraint} we identify the explicit relational Hamiltonian evolving along the clock $T$:
\begin{equation}
\label{eq:Relational_Hamiltonian_Supermetric}
    H_r(\phi^A, p_A) = \pm \sqrt{\gamma^{AB}(\phi) p_A p_B} \,.
\end{equation}
Because $p_c$ appears quadratically, extracting $H_r$ requires choosing a sign for the square root. Physically, this $\pm$ branch dictates whether the chosen clock ticks forward or backward with respect to the underlying gauge evolution, thereby establishing a specific arrow of relational time for the reduced dynamics (e.g., distinguishing between an expanding or a contracting universe in cosmology).

With this identification, the internal variable $T$ acts as the evolution parameter, and $H_r$ acts as the true generator of physical (relational) evolution. The relational dynamics on this reduced phase space is then simply given by the standard Hamilton equations with respect to the clock $T$:
\begin{equation}
    \frac{\d \phi^A}{\d T} = \{\phi^A, H_r\}_{\text{red}} \,, \quad  \quad \frac{\d p_A}{\d T} = \{p_A, H_r\}_{\text{red}} \,,
\end{equation}
where $\{\cdot,\cdot\}_{\text{red}}$ denotes the Poisson bracket evaluated on the $2(d-1)$-dimensional non-clock sector. Note that it is precisely this physical Hamiltonian $H_r$ and these reduced brackets that directly feed into the relational observable $Q_{f_r, T}(t)$ introduced in Eq.~\eqref{eq:Relational_observables_simplified}.

Having established the reduced phase-space dynamics, it is instructive to examine why this physical framework is typically constructed within the Hamiltonian formalism rather than relying on a reduced Lagrangian. Using the relational velocities $\phi'^A \coloneqq \d\phi^A/\d T$, one can write the fully deparametrised relational action in first-order form as
\begin{equation}
    \label{eq:Reduced_Action_Velocities}
    S_{\text{red}} = \int_{\mathbb{R}} \d T \, \left( p_A \phi'^A - H_r(\phi^A, p_A) \right) \,.
\end{equation}
While this is a perfectly well-defined action for an unconstrained Hamiltonian system, one cannot simply replace the momenta in terms of velocities or perform a standard Legendre transform to obtain a second-order relational Lagrangian $L_r(\phi^A, \phi'^A)$. Because $H_r$ in Eq.~\eqref{eq:Relational_Hamiltonian_Supermetric} is a square root of a quadratic form, it is a homogeneous function of degree one in the momenta. By Euler's theorem on homogeneous functions, this implies $p_A \, \partial H_r / \partial p_A = H_r$, and since Hamilton's equations read $\phi'^A = \partial H_r / \partial p_A$, the corresponding relational Lagrangian $L_r = p_A \phi'^A - H_r$ evaluates to zero on-shell \cite{Gaugebook}. This degenerate Lagrangian behaviour highlights why the purely relational dynamics of such systems are best handled using the Hamiltonian phase-space machinery.

To circumvent this degeneracy when a Lagrangian description based on the supermetric is required in practice, one can return to the unreduced configuration space and explicitly evaluate the clock-adapted lapse \eqref{eq:u_multipl} to obtain a gauge-fixed description. In particular this allows to write a non-degenerate, \textit{gauge-fixed} action, which can provide a useful structural framework for the study of symmetries based on minisuperspace metric. In doing so, we retain the full $2d$-dimensional phase space but demand that the time parameter ticks synchronously with the physical clock (such that $\dot{T} = 1$). Then, using the constraint \eqref{eq:diagonal_constraint}, the Poisson bracket $\{T, h\} = \{\phi^c, h\} = g^{cc}p_c$ dynamically locks the lapse function to the specific expression
\begin{equation}
    \label{eq:adapted_lapse_metric}
    N = \frac{1}{g^{cc}p_c} = \frac{g_{cc}}{p_c} \,,
\end{equation}
where again $g^{cc}=1/g_{cc}$ by diagonality. While this specific choice of $N$ merely fixes a coordinate frame along the gauge orbit without reducing the system's dimensionality (and hence without properly deparametrising), it is often a highly convenient step for practical purposes. Substituting the adapted lapse \eqref{eq:adapted_lapse_metric} into the unreduced action one finds 
\begin{equation}
    S_{\text{gf}} = \tfrac{1}{2} \int \d T \, p_c\left(1 - \gamma_{AB}\, \phi'^A \phi'^B \right)\,.
\end{equation}
Here $p_c$ is to be viewed as an independent auxiliary field: its variation enforces
$\gamma_{AB}\,\phi'^A \phi'^B = 1$, the Lagrangian counterpart of the mass-shell relation \eqref{eq:massivegeo}, while variation of the $\phi^A$ yields the corresponding massive geodesic equations. Note that it is straightforward to properly deparametrise the dynamics obtained from such a gauge-fixed action at the level of equations of motion, since the gauge time parameter can be directly eliminated in favour of the physical degree of freedom acting as the relational clock (the condition $\dot T = 1$ integrates to $T = \phi^c + \text{const}$). Importantly, since Dirac observables are strictly gauge-invariant, the final physical evolution remains entirely independent of this (or any other) choice of lapse.

Crucially, however, selecting a specific physical variable to serve as a clock carries important consequences for the manifest symmetry structure of the theory, whether one proceeds via deparametrisation or via the gauge-fixed action. While the unreduced constraint \eqref{eq:Hamiltonian_Constraint_Supermetric} describes null paths on the full configuration space (so that its natural dynamical symmetries are governed by the conformal isometries of $g_{ab}$), forcing a clock variable into a preferential role breaks this manifest conformal covariance. In particular, on the constraint surface the clock momentum satisfies the mass-shell relation \eqref{eq:massivegeo}, so that the null-cone structure of the unreduced dynamics is traded for a massive one on the non-clock sector. When the supermetric does not depend on the chosen clock, $p_c$ is a constant of motion and the projected trajectories are precisely {massive} geodesics of the reduced metric $\gamma_{AB}$, with $|p_c|$ playing the role of an effective mass. This restricts (relational) time-independent conserved charges to the standard Killing vectors (isometries) of the reduced metric $\gamma_{AB}$. The general conclusion is that conformal transformations cease to map solutions onto solutions, and the full symmetry algebra of the system is truncated into a smaller, clock-dependent subalgebra. This observation frames the Eisenhart--Duval lift \cite{ED,ED2,ED3}---which we adopt in Section~\ref{sec:Lift}---as a general diagnostic tool. Indeed, if one is given \textit{a priori} a system governed by a square-root Hamiltonian $H \propto \sqrt{\gamma^{AB}p_A p_B}$ (cf.~\eqref{eq:Relational_Hamiltonian_Supermetric}) without knowing whether it originally emerged from a time-reparametrisation invariant theory, standard phase-space methods would remain structurally blind to its hidden conformal symmetries, uncovering only the restricted set of exact isometries.

\section{Minisuperspace symmetries of FLRW cosmology with $n$ scalar fields}
\label{sec:FLRW}

In this section we ground the relational framework described above into a concrete physical system: a flat Friedmann--Lemaître--Robertson--Walker (FLRW) universe minimally coupled to $n$ free, massless scalar fields, extending the single-field case of~\cite{EteraDaniele}. This model provides an ideal testing ground for symmetry analysis. Because its minisuperspace potential vanishes identically, the classical trajectories map directly to the null geodesics of the underlying field-space metric. We first explicitly derive the relational dynamics using both a matter and a geometric internal clock, and then analyse the system's minisuperspace symmetry structure---specifically through its Conformal Killing Vectors (CKVs)---focusing on both coordinate-time and fully gauge-invariant relational formulations.

\subsection{Relational dynamics of flat FLRW cosmology}
\label{subs:FLRW_dynamics}

We start from the full Einstein--Hilbert action minimally coupled to $n$ massless scalar fields $\chi^1,\dots,\chi^n$:
\begin{equation}
    \label{eq:Full_action}
     S[g,\chi_i]= \int
     {\rm{d}}^4 x \sqrt{\vert g \vert} \left[  \frac{R}{16 \pi G} - \frac{1}{2} g^{\mu \nu} \sum_{i=1}^{n} \partial_\mu \chi^i \, \partial_\nu \chi^i  \right].
\end{equation}
We consider a flat FLRW metric, which in coordinates adapted to its spacetime symmetries (homogeneity and isotropy) reads
\begin{equation}
    \label{eq:FLRW_line_element}
    \d s^2=-N^2(t) \d t^2+a^2(t) \delta_{ij} \,{\rm{d}}x^ i {\rm{d}}x^j \,,
\end{equation}
where $N(t)$ is the lapse function and $a(t)$ the scale factor. The symmetry reduction in the matter sector is realised by having the matter fields be independent of spatial coordinates $x^i$. Following the notation of \cite{EteraDaniele}, we can then write the symmetry reduced Einstein--Hilbert action \eqref{eq:Full_action} as
\begin{equation}
\label{eq:Symmetry_reduced_action}
    S = c\,l_p\int_{\mathbb{R}} \frac{{\rm{d}} t}{2 N}\left[ l_p^2 z^2 \sum_{i=1}^n (\dot \chi^i)^2 -4 \dot z^2\right],
\end{equation}
where the dot denotes derivatives with respect to the gauge time parameter $t$, we have defined $z \coloneqq a^{3/2}$, and we have integrated over a fiducial cell of volume $V_0$, defining as well $l_p \coloneqq \sqrt{12 \pi G}$ and $c \coloneqq V_0/l_p^3$. 

Since diffeomorphism symmetry is reduced to time-reparametrisation
invariance, the symmetry-reduced action \eqref{eq:Symmetry_reduced_action} conforms exactly to the general supermetric framework described in Section \ref{subsec:Supermetric} with a vanishing potential. The gravitational system can be seen as the motion of a point particle on minisuperspace, i.e., the $(n+1)$-dimensional configuration space $\mathcal{Q}$. This is spanned by the coordinates $\phi^a = (z, \chi^1, \dots, \chi^n)$ with $a\in\{0,1,\dots,n\}$, and is equipped with a purely diagonal minisuperspace metric $g_{ab}$ with
\begin{equation}
    \label{eq:FLRW_inverse_metric}
    g_{zz} = -{4 c l_p} \,, \quad  \quad g_{ij} = {c l_p^3 z^2} \delta_{ij} \,.
\end{equation}
The canonical momenta conjugate to the configuration variables are given by
\begin{equation}
    p_z = -\frac{4cl_p}{N}  \dot{z}\,, \qquad p_i = \frac{c l_p^3 z^2}{N} \dot{\chi}^i\,,
\end{equation}
with canonical Poisson brackets $  \{z,p_z\}=\{\chi^i,p_i\}=1$. As we have seen, variation with respect to the lapse $N$ enforces the Hamiltonian constraint
\begin{equation}\label{eq:h}
h=
\frac{1}{2 c l_p^3 z^2} \sum_{i} p_i^2 - \frac{p_z^2}{8 c l_p}\approx 0\,,
\end{equation}
defining the constraint surface $\mathcal{C}_H$. This is a $(2n+1)$-dimensional conic subset of the cotangent bundle over minisuperspace, fibred by $n$-dimensional double cones. For each fixed value of $(z,\chi^i)$, the fibre cone is defined by
\begin{equation}
    \left(\frac{l_p z p_z}{2}\right)^2=\sum_i p_i^2 \,.
\label{eq:ConstrainSurfaceFLRW}
\end{equation}
The vertex of each double-cone fibre is located at $p_z=p_i=0$, and the half-opening angle is given by $\tan\theta={l_p z}/{2}$. Characterising $\mathcal{C}_H$ through this conic geometry is physically relevant: it identifies
the physical phase space with the local null (causal) structure of the minisuperspace. Since
conformal transformations preserve null cones, the classical trajectories running along these
cones are insensitive to conformal rescalings of the supermetric. This observation anticipates
why conformal Killing vectors provide the natural geometric tool to uncover the symmetries of
the system, as we exploit in Section~\ref{sub:FLRW_Symmetries}. 

Solving the canonical equations of motion with respect to the coordinate time $t$ straightforwardly yields the standard Friedmann dynamics. Rather than focusing on these well-established coordinate-dependent dynamics, we proceed directly to the construction of a fully gauge-invariant description through the relational framework depicted in the previous section, directly applying the deparametrisation procedure. Our general framework accommodates selecting either a matter field or the geometric scale factor, each yielding a distinct reduced phase-space structure that parametrises the same underlying set of fully gauge-invariant Dirac observables in different ways.

\paragraph{Matter Clock.} If we select one of the scalar fields as the internal clock, e.g., $T \coloneqq \chi^n$, the physical non-clock sector is spanned by $(z, \chi^1, \dots, \chi^{n-1})$. From Eq.~\eqref{eq:gamma_definition} one finds $\gamma^{AB} = \text{diag}(\frac{1}{4}l_p^2 z^2, -1, \dots, -1)$, which dictates the (Lorentzian) geometry of the physical relational space. The relational Hamiltonian \eqref{eq:Relational_Hamiltonian_Supermetric} driving the physical evolution takes the form
\begin{equation}
    \label{eq:FLRW_Hr_Matter}
    H_{r}^{(\text{matter})} = \pm \sqrt{ \frac{l_p^2 z^2}{4} p_z^2 - \sum_{i=1}^{n-1} p_i^2 } \,.
\end{equation}
Solving the corresponding Hamilton equations yields the purely relational, gauge-invariant dynamics. Identifying the constant of motion $P_z \coloneqq z p_z$, the explicit physical trajectories parametrised by the matter clock relative to an arbitrary clock initial value $\chi^n_0$ are given by:
\begin{equation}
\label{eq:RelationalObsFLRW_Matter}
\begin{split}
    z(\chi^n) &= z_0 \, \exp\left[ \frac{l_p^2 P_z}{4 H_r^{(\text{matter})}} (\chi^n - \chi^n_0) \right] \,, \\
    p_z(\chi^n) &= p_{z,0}\, \exp\left[ -\frac{l_p^2 P_z}{4 H_r^{(\text{matter})}} (\chi^n - \chi^n_0) \right] \,, \\
    \chi^i(\chi^n) &= -\frac{p_i}{H_r^{(\text{matter})}} (\chi^n - \chi^n_0) + \chi^i_0 \,, \\
    p_i &= p_{i,0}= \text{const} \,,
\end{split}
\end{equation}
where $z_0 \equiv z(\chi^n_0)$, $p_{z,0} \equiv p_z(\chi^n_0)$, $\chi^i_0 \equiv \chi^i(\chi^n_0)$ and $p_{i,0} \equiv p_i(\chi^n_0)$ are initial conditions.

\paragraph{Geometric Clock.} Alternatively, if we use the geometric variable as the clock, $T \coloneqq z$, the physical sector consists purely of the $n$ matter fields. The reduced metric on this subspace is Euclidean, given by $\gamma^{ij} = \frac{4}{l_p^2 z^2} \delta^{ij}$, and the relational Hamiltonian is
\begin{equation}
    \label{eq:FLRW_Hr_Geom}
    H_{r}^{(\text{geom})} = \pm \frac{2}{l_p z} \sqrt{ \sum_{i=1}^n p_i^2 } \,.
\end{equation}
Note that while the clock $\chi^n$ is cyclic (so that \eqref{eq:FLRW_Hr_Matter} is a constant of motion), $z$ explicitly appears in the minisuperspace metric and hence the relational Hamiltonian $H_r^{(\text{geom})}$ is time dependent. Importantly, despite this explicit clock dependence, the dynamics remains straightforward to solve as the clock variable $z$ completely factors out in \eqref{eq:FLRW_Hr_Geom}. Solving Hamilton's equations gives the relational evolution of the matter fields with respect to the scale factor. As expected, these solutions perfectly mirror the algebraic inverses of the matter-clock trajectories \eqref{eq:RelationalObsFLRW_Matter}. Using $h \approx 0$ (specifically \eqref{eq:ConstrainSurfaceFLRW}), we can substitute the total matter momentum using $\sum_{i=1}^n p_i^2 = l_p^2 P_z^2 / 4$, where $P_z$ is defined below \eqref{eq:FLRW_Hr_Matter}, and obtain the explicit trajectories:
\begin{equation}
\label{eq:RelationalObsFLRW_Geom}
\begin{split}
    \chi^i(z) &= \chi^i_0 -\frac{4 p_i}{l_p^2 P_z} \ln\left(\frac{z}{z_0}\right) \,, \\
    p_i &= p_{i,0}= \text{const} \,,
\end{split}
\end{equation}
where $\chi^i_0 \equiv \chi^i(z_0)$ and $p_{i,0} \equiv p_i(z_0)$. Note that the momentum conjugate to the clock is identically fixed by the deparametrisation constraint as $p_z = -H_r^{(\text{geom})} = P_z/z$, fully consistent with the matter clock formulation.

These explicit trajectories provide the concrete realisations of the complete relational observables formalised in Eq.~\eqref{eq:Relational_observables} and \eqref{eq:Relational_observables_simplified}. Physically, equations \eqref{eq:RelationalObsFLRW_Matter} and \eqref{eq:RelationalObsFLRW_Geom} predict the exact value of any physical non-clock variable at the precise ``instant'' the designated internal clock reads a given value. By construction, these relational functions constitute weak Dirac observables; they Poisson-commute with the Hamiltonian constraint $h$ on the constraint surface $\mathcal{C}_H$. Consequently, they remain strictly invariant under the time-reparametrisation gauge flow. This demonstrates the ultimate utility of the deparametrisation procedure: it strips away the unphysical coordinate scaffolding while preserving a clear, purely relational description of cosmological evolution.\footnote{The role of the underlying manifold on which fields were supported before deparametrisation remains that of providing only topological constraints on the space of allowed field configurations.}

\subsection{Hidden symmetries via conformal Killing vectors}
\label{sub:FLRW_Symmetries}

As established in the previous section, the dynamics of our flat FLRW universe coupled to $n$ scalar fields is entirely encoded in the supermetric geometry of its configuration space. Specifically, the classical trajectories
are null geodesics of the $(n+1)$-dimensional minisuperspace metric $g_{ab}$, a
structure preserved by conformal transformations. The conformal isometries of $g_{ab}$ therefore map onto continuous dynamical symmetries of the cosmological system, which we now examine.

To systematically extract the associated conserved quantities, we evaluate the conformal Killing vector fields (CKVs) of the FLRW supermetric. A vector field $\xi = \xi^a(\phi)\,\partial_a=\xi^a(z, \chi^i)\partial_a$ on minisuperspace is a CKV if and only if it satisfies
\begin{equation}
    \label{eq:CKV_definition}
    {\mathcal{L}}_\xi\, g_{ab}= \varphi\, g_{ab} \,,
\end{equation}
where $\varphi(\phi)=\varphi(z, \chi^i)$ is the conformal factor associated with $\xi$. By Noether's theorem, every such CKV generates a conserved charge $Q_\xi$ that is linear in the momenta, defined explicitly as:
\begin{equation}
    \label{eq:Qxi}
    Q_\xi \coloneqq\sum_a \xi^a\,p_a= \xi^z p_z + \sum_{i=1}^n \xi^i p_i \,.
\end{equation}
The on-shell conservation of this charge is verified by evaluating the Poisson bracket with $H = N h$:
\begin{equation}\label{eq:CKV_conservation}
    \{Q_\xi, N \,h\} = 
    \frac{N}{2} p^a p^b {\mathcal{L}}_\xi g_{ab} = 
    N\, \varphi \,h \approx 0 \,,
\end{equation}
where we have used the constraint condition $h \approx 0$ (explicitly, \eqref{eq:h}) and the CKV definition \eqref{eq:CKV_definition}. Consequently, $Q_\xi$ constitutes a weak Dirac observable that generates a physical symmetry on the constraint surface $\mathcal{C}_{H}$. 

A standard result in differential geometry dictates that the maximum number of linearly independent CKVs on a $d$-dimensional manifold is given by $(d+1)(d+2)/2$ for $d \geq 3$,\footnote{We leave the discussion of the case with one scalar field (so that $d=2$) for later.} with a manifold admitting this maximal bound if and only if it is conformally flat. As we will see, our $(n+1)$-dimensional FLRW minisuperspace precisely realises this upper bound. Because the map \eqref{eq:Qxi} from the vector space of CKVs to the space of phase-space functions is an injective linear homomorphism, the linear independence of these geometric vector fields directly guarantees the linear independence of their corresponding Noether charges. Furthermore, their Poisson brackets mirror the Lie algebra of the vector fields since $\{\xi^a p_a, \zeta^b p_b\} =  -Q_{[\xi, \zeta]}$, meaning that the conserved charges form a closed Poisson algebra isomorphic to the conformal group, as we will show below.

Crucially, we will construct charges that possess no explicit dependence on the unphysical coordinate time $t$, thus constituting a set of gauge-invariant weak Dirac observables on the constraint surface $\mathcal{C}_H$. However, because they are expressed in terms of the internal minisuperspace degrees of freedom, designating one of these variables as a relational clock (as explained in the previous section) naturally translates any charge containing that variable into an explicitly (relational) time-dependent symmetry relative to that internal, physical clock (see \eqref{eq:Qr_Conserved_wrt_Any} and discussion thereunder). One might still question whether there exist broader, explicitly $t$-dependent symmetries that this standard configuration-space approach misses. To answer this, we will later employ the Eisenhart--Duval lift---a higher-dimensional geometric framework that naturally incorporates $t$-evolution \cite{ED,ED2,ED3}. We will show that the explicitly $t$-dependent charges obtained in this way are gauge-dependent objects, intertwining physical symmetries with time reparametrisations.

To explicitly construct the symmetry structure of our minisuperspace, we perform a direct integration of the CKV equations \eqref{eq:CKV_definition} over our specific minisuperspace configuration. The most general solution \eqref{eq:Qxi} yields a total charge
$Q = \xi^z p_z + \sum_i \xi^i p_i$ parametrised by a set of arbitrary integration constants
($c_0$, $c_i$, $c_{ij}$, $\lambda$, $b_i$, $\beta$ and $\beta_i$):
\begin{equation}
\label{eq:Q_general}
\begin{aligned}
    Q  ={}& c_0 (z p_z) + \sum_{i} c_i p_i + \sum_{i < j} c_{ij} (\chi^i p_j - \chi^j p_i) + \lambda \left( z p_z \ln z + \sum_{i=1}^n \chi^i p_i \right) \\
    &+ \sum_{i=1}^n b_i \left( \frac{l_p}{2} z p_z \chi^i + \frac{2}{l_p} p_i \ln z \right) + \beta \left( \frac{l_p z p_z}{4} \left[ \sum_{i=1}^n (\chi^i)^2 + \frac{4}{l_p^2} (\ln z)^2 \right] + \frac{2 \ln z}{l_p} \sum_{i=1}^n p_i \chi^i \right) \\
    &    + \sum_{i=1}^n \beta_i \left( \chi^i \sum_{j=1}^n \chi^j p_j
      - \frac{p_i}{2} \sum_{j=1}^n (\chi^j)^2
      + z\, p_z\, \chi^i \ln z + \frac{2 p_i}{l_p^2} (\ln z)^2 \right)\,.
\end{aligned}
\end{equation}
Isolating the independent terms associated with each integration constant, we extract the
complete set of $(n+2)(n+3)/2$ conserved charges:
\begin{equation}
\label{eq:charges_physical}
\begin{aligned}
    P_i &= p_i \,, \\
    J_{ij} &= \chi^i p_j - \chi^j p_i \,, \\
    P_z &= z p_z \,, \\
    B_i &= \frac{l_p}{2} z p_z \chi^i + \frac{2}{l_p} p_i  \ln z \,, \\
    D &= z\, p_z\,\ln z + \sum_{i=1}^n \chi^i\,p_i \,, \\
    K_z &= \frac{l_p\,z\,p_z}{4}\left[ \sum_{i=1}^n( \chi^i)^2 + \frac{4}{l_p^2}  (\ln z)^2 \right]+ \frac{2 \,\ln z}{l_p} \sum_{i=1}^n p_i\,\chi^i \,,\\
     K_i &= \chi^i\,\sum_{j=1}^n p_j \chi^j - \frac{p_i}{2} \sum_{j=1}^n (\chi^j)^2 + z\,p_z \chi^i\,\ln z + \frac{2 p_i}{l_p^2} (\ln z)^2 \,.
\end{aligned}
\end{equation}
Let us note that the $(n+2)(n+3)/2$ charges vastly outnumber the phase-space dimensions, so that they cannot all be functionally independent. Moreover, being Dirac observables ($P_i$ and $J_{ij}$ strongly, the remaining charges only weakly, see \eqref{eq:Dirac_observable} and
\eqref{eq:strong}), they are constant along gauge orbits, which reduces the number of independent values further. This redundancy can be put to work: as explained in Appendix~\ref{app:reconstruction}, the relational trajectories \eqref{eq:RelationalObsFLRW_Matter} and \eqref{eq:RelationalObsFLRW_Geom} can be reconstructed purely algebraically from the conserved charges, without integrating any equation of motion.

To determine the algebraic structure of these constants of motion, it is convenient to map the minisuperspace onto canonical conformal coordinates:
\begin{equation}\label{eq:XPmunu}
    X^\mu = \left( -\frac{2}{l_p} \ln z ,\, \chi^i \right) \,, \qquad P_\mu = \left( -\frac{l_p}{2} \,z\,p_z ,\, p_i \right) \,,
\end{equation}
where $\mu, \nu = 0, 1, \dots, n$. In this coordinate system, the metric indices are raised and lowered using the $(n+1)$-dimensional Minkowski metric $\eta_{\mu \nu} = \text{diag}(-1, +1, \dots, +1)$. This coordinate choice reveals that the supermetric \eqref{eq:FLRW_inverse_metric} is conformally flat, as its line element collapses into the simple form:
\begin{equation}\label{eq:confflatX}
    \d s^2 = c\,l_p^3 e^{- l_p X^0} \left[ -(\d X^0)^2 + \sum_{i=1}^n (\d \chi^i)^2 \right] \,.
\end{equation}
Equivalently, translating the Hamiltonian constraint \eqref{eq:h} into these canonical coordinates yields:
\begin{equation}
h =  \frac{e^{l_p X^0}}{2 c l_p^3} \eta^{\mu\nu} P_\mu P_\nu = \frac{e^{l_p X^0}}{2 c l_p^3} \left(-P_0^2 + \sum_i p_i^2\right) \approx 0 \,,
\end{equation}
showing that the constraint is just the null mass-shell condition in minisuperspace.

Exploiting these coordinates, the most general CKV $\xi = \xi^\mu \partial_\mu$ can be written compactly in terms of arbitrary integration constants as the standard conformal vector field of a flat space:
\begin{equation}
\xi^\mu(X) = a^\mu + \omega^\mu{}_{\nu} X^\nu + \lambda X^\mu + 2(k_\nu X^\nu)X^\mu - (X_\nu X^\nu)k^\mu \,,
\end{equation}
where the parameters $a^\mu$ govern translations, $\omega_{\mu\nu} = -\omega_{\nu\mu}$ govern rotations and boosts, $\lambda$ governs dilatations, and $k^\mu$ govern special conformal transformations.\footnote{Mapping these generators back to phase space via $Q = \xi^\mu P_\mu$, the parameters relate to the integration constants of Eq.~\eqref{eq:Q_general} as $a^0 = -\frac{2}{l_p}c_0$, $a^i = c_i$, $\omega^{ij} = -c_{ij}$, $\omega^{0i} = -b_i$, $k^i = \tfrac{1}{2}\beta_i$, and $k^0 = \frac{1}{2}\beta$.} With this notation, the full set of conserved quantities listed in Eq.~\eqref{eq:charges_physical} can be rewritten in a covariant way as
\begin{equation}\label{eq:MDKcharges}
\begin{aligned}
    M_{\mu\nu} &= X_\mu P_\nu - X_\nu P_\mu \,, \\
    D        &= X_\mu P^\mu \,, \\
    K_\mu    &= X_\mu (P_\nu X^\nu) - \frac{1}{2} P_\mu (X_\nu X^\nu) \,.
\end{aligned}
\end{equation}
Importantly, their Poisson brackets yield the completely closed algebraic structure:
\begin{equation}
\label{eq:conf_algebra}
\begin{split}
    \{ D , K_\mu \} &= -K_\mu \,, \\
    \{ D, P_\mu \} &= P_\mu \,, \\
    \{K_\mu,P_\nu\} &= \eta_{\mu \nu} D + M_{\mu \nu} \,, \\
        \{K_\mu , M_{\nu \rho} \} &= \eta_{\mu \rho} K_\nu - \eta_{\mu \nu} K_\rho \,, \\
    \{P_\mu , M_{\nu \rho} \} &= \eta_{\mu \rho} P_\nu - \eta_{\mu \nu} P_\rho \,, \\
    \{M_{\mu \nu} , M_{\rho \sigma} \} &= \eta_{\mu \rho} M_{\nu \sigma} + \eta_{\nu \sigma} M_{\mu \rho} - \eta_{\mu \sigma} M_{\nu \rho} - \eta_{\nu \rho} M_{\mu \sigma} \,.
\end{split}
\end{equation}
This system explicitly identifies the underlying dynamical symmetry algebra as the conformal Lie algebra $\mathfrak{conf}(n,1)\simeq \mathfrak{so}(n+1,2)$, possessing the maximal dimension of ${(n+2)(n+3)}/{2}$. Physically, the generators correspond to \textit{minisuperspace} translations $P_\mu$, spatial rotations $M_{ij}$, dilatations $D$, Lorentz boosts $M_{0i}$ (equivalent to the charges $B_i$), and special conformal transformations $K_\mu$ (encompassing both $K_z$ and $K_i$).\footnote{This construction straightforwardly generalises to models with $k$ ghost-like scalar fields, characterised by kinetic terms with opposite sign. The underlying geometry remains conformally flat, but the metric has signature $\eta^{(\varepsilon)}_{\mu\nu} = \text{diag}(-1, \varepsilon_1, \dots, \varepsilon_n)$ where $\varepsilon_i = \pm 1$, featuring $1+k$ timelike and $n-k$ spacelike directions. The CKV derivation proceeds identically by replacing Euclidean contractions with their $\varepsilon$-weighted counterparts---e.g., $\sum (\chi^i)^2 \to \sum \varepsilon_i (\chi^i)^2$. Consequently, the dynamical algebra naturally generalises to $\mathfrak{conf}(n-k, 1+k) \simeq \mathfrak{so}(n-k+1, k+2)$, with the minor physical distinction that some $M_{ij}$ generators now induce boosts depending on the signs of the fields they mix.}

Crucially, the charges obtained above are weak Dirac observables (see \eqref{eq:Dirac_observable}): their conservation is not tied to a particular choice of clock and they can be viewed as evolving constants of motion with respect to \emph{any} admissible relational time, as in Eq.~\eqref{eq:Qr_Conserved_wrt_Any}. Then, for the specific (free) system at hand, there is in principle no need to repeat the symmetry analysis after deparametrisation. However, more realistic models incorporating spatial curvature or scalar field potentials break the underlying null geodesic structure of the unreduced dynamics (needed to map dynamical symmetries to conformal isometries, see discussion above Eq.~\eqref{eq:CKV_definition}). In such cases, looking for conformal Killing vectors is not a viable strategy, and one needs alternative methods for symmetry extraction.  

Although the symmetry analysis is already complete for the FLRW universe with $n$ scalars (having found the maximal number of independent charges \eqref{eq:MDKcharges}), this model provides an explicit illustration of the general mechanism discussed at the end of Section~\ref{subsec:Supermetric}. We recall that once a clock is singled out (whether via the reduced relational Hamiltonian or the gauge-fixed action), the clock momentum obeys the mass-shell relation \eqref{eq:massivegeo} on the constraint surface, and the null trajectories of the unreduced theory are traded for \emph{massive} geodesics of the reduced metric $\gamma_{AB}$ (see \eqref{eq:gamma_definition}). This implies that the symmetries are reduced to the isometries of $\gamma_{AB}$. For the two clocks of Section~\ref{subs:FLRW_dynamics} these isometries can be read off directly from the reduced metrics computed there: the matter-clock metric $\gamma^{AB}=\mathrm{diag}(\tfrac14 l_p^2z^2,-1,\dots,-1)$ is flat and Lorentzian, so its isometries form the Poincaré algebra $\mathfrak{iso}(n-1,1)$, while the geometric-clock metric $\gamma^{ij}=\tfrac{4}{l_p^2z^2}\,\delta^{ij}$ is Euclidean up to a clock-dependent scale, leaving the Euclidean algebra $\mathfrak{iso}(n)$. In either case the algebra $\mathfrak{conf}(n,1)$ collapses to a small, clock-dependent subalgebra, even though its charges remain perfectly valid conserved Dirac observables. Since the full conformal algebra consists of perfectly valid conserved Dirac observables, it is clear that the standard reduced phase-space method systematically misses physically relevant, relational-time-dependent charges. To see how these missing charges can systematically be recovered from a relational perspective, one requires a higher-dimensional framework capable of geometrising the physical clock time itself---a task we turn to in the next section via the Eisenhart--Duval lift. Before doing that, let us comment on the special case of a single scalar field.

\subsubsection*{The special case of a single scalar field ($n=1$)}

For an FLRW universe with a single (free and massless) scalar field, the minisuperspace is two-dimensional, with coordinates $(z,\chi)$ and metric $g_{ab}=\text{diag} \left(-4 cl_p,cl_p^{3}z^{2}\right)$ (cf.~\eqref{eq:FLRW_inverse_metric}). This requires a separate treatment due to the well-known fact that the algebra of local conformal transformations is infinite-dimensional in two dimensions. Indeed, the CKV condition $\mathcal{L}_\xi g_{ab} = \varphi g_{ab}$ simplifies to a $(1+1)$-dimensional wave equation \cite{DiFrancesco:1997nk}, where the general solutions decouple into arbitrary left- and right-moving modes along the null coordinates $\sigma = \chi - X^0$ and $\rho=\chi+X^0$ (where $X^0$ is defined in \eqref{eq:XPmunu}). The phase-space charges correspondingly split into two independent sectors, each realising a \textit{Witt algebra}. The full dynamical symmetry algebra is therefore the infinite-dimensional direct sum $\mathfrak{Witt}\oplus \mathfrak{Witt}$, as explicitly shown in \cite{EteraDaniele}.

This behaviour highlights an important geometric distinction between the single-field and multi-field scenarios. For models with $n \ge 2$, the minisuperspace is conformally flat but genuinely curved (see \eqref{eq:confflatX}) and its strict isometries---obtained by setting $\varphi=0$ in \eqref{eq:CKV_definition}---form only the purely spatial Euclidean algebra $\mathfrak{iso}(n)$. In contrast, the $n=1$ minisuperspace possesses identically vanishing curvature and its proper isometries manifest as the full three-dimensional Poincaré algebra $\mathfrak{iso}(1,1)$. This is easy to see by transitioning to the light-cone variables $x^\pm: = {z}\,e^{\pm \frac{l_p}{2}\chi}$ (with $x^+=e^{\frac{l_p}{2}\sigma}$ and $x^-=e^{-\frac{l_p}{2}\rho}$), as the charges organise naturally into light-cone momenta and a Lorentz boost \cite{EteraDaniele}
\begin{equation}\label{eq:n1NO}
P_\pm=e^{\mp\frac{l_p}{2}\chi}\left(\frac{p_z}{2}\pm\frac{p_\chi}{l_p z}\right)\,, \qquad J=\frac{p_\chi}{l_p} \,.
\end{equation}
This makes it explicitly clear that the original minisuperspace variables $(z, \chi)$ simply serve as hyperbolic polar coordinates on a strictly flat two-dimensional Minkowski space.

Ultimately, since $\mathfrak{Witt}\oplus \mathfrak{Witt}\cong\mathfrak{conf}(1,1)$, our general framework is completely consistent with this previous result \cite{EteraDaniele} and extends it to any number of fields: the underlying dynamical symmetry remains fundamentally conformal for arbitrary $n$, with the infinite-dimensional structure being purely an intrinsic property of two dimensions.

\section{Geometrising (relational) time with the Eisenhart--Duval lift}\label{sec:Lift}

In the previous section we identified the symmetries of the minisuperspace of an FLRW universe with $n$ massless scalar fields by analysing its conformal Killing vectors, revealing the rich conformal algebra $\mathfrak{conf}(n,1)\simeq\mathfrak{so}(n+1,2)$. While this algebra acts globally on configuration space, one may wonder how dynamics affects this result. As emphasised in Section \ref{sec:Constrained}, in a reparametrisation-invariant theory the concept of dynamics is subtle: evolution can be defined either with respect to a coordinate time parameter $t$ or relative to an internal physical degree of freedom. While the analysis carried out in Section \ref{sec:FLRW} is unable to systematically derive coordinate-time dependence, the conserved charges we have found describe relations between minisuperspace degrees of freedom, and can be understood as \textit{relationally} clock-time-dependent and thus physical. 

Importantly, we have seen that, in a deparametrised description, the momentum of the chosen clock introduces a square-root physical Hamiltonian and acts as an effective mass (cf.~\eqref{eq:massivegeo} and \eqref{eq:Relational_Hamiltonian_Supermetric}). This results in the symmetry algebra collapsing to a clock-dependent isometry subalgebra, while the remaining
clock-time-dependent charges drop out of sight. These charges are not lost: they
are relational observables whose value depends on the clock reading, and
capturing them systematically requires a framework in which (possibly relational) time dependence is incorporated into the symmetry generators themselves. A standard tool for geometrising explicit time dependence in classical mechanics is the Eisenhart--Duval (ED) lift \cite{ED}, briefly reviewed in Appendix \ref{app:ED}. This embeds the system into an extended manifold where dynamical trajectories correspond to null geodesics (just like in Section \ref{sub:FLRW_Symmetries}) of a Lorentzian metric $G$ in two additional dimensions \cite{ED2,ED3}. Crucially, the conserved charges identified within this enlarged space are not detached from the physical theory; they can be systematically mapped back to the constants of motion of the original mechanical system via well-defined projection rules. 

For the free FLRW model with $n$ scalar fields, the minisuperspace trajectories are already null since the Hamiltonian constraint is quadratic. However, the
conceptual and technical necessity of the lift becomes fully apparent
when one turns to more realistic cosmological models, which can in general include curvature or matter potentials. Such potentials, just like the effective mass of the relational description, force the physical trajectories \textit{off} the minisuperspace null cone. The extra dimensions of the ED lift absorb precisely these potentials and effective masses into the lifted metric $G$, restoring a higher-dimensional null-geodesic formulation.

In this section we apply the ED lift to our FLRW model, with the aim of clarifying the delicate interplay between the notion of \textit{relational} constants of motion and the choice of gauge. We will see that the symmetries of the
lifted system are gauge-dependent, and  controlled by the curvature of the
gauge-fixed metric $\tilde g=g/N$ (cf.~Appendix~\ref{app:ED}). We will then extract the gauge-invariant (relational) content by deparametrisation, showing that the relational charges always lie within $\mathfrak{conf}(n,1)$, which they exhaust in a specific gauge (associated with the chosen clock).

\subsection{The lift in a family of gauges}\label{sec:lift_family}

 When implementing the ED lift one needs to solve the generalised conformal Killing equations
\begin{equation}\label{eq:XiG}
    \mathcal{L}_\Xi \, G_{AB} = \Omega \, G_{AB} \,,
\end{equation}
analogous to \eqref{eq:CKV_definition} (see Appendix \ref{app:ED} for details). While one can formally perform the ED lift procedure without spelling out the specific form of the lapse, solving \eqref{eq:XiG} in closed form requires an explicit functional form for $N$. Rather than restricting our analysis to a single gauge (as commonly done in previous work \cite{EteraDaniele}), we use a class of lapses
\begin{equation}\label{eq:OurLapse}
    N(z) = \alpha \, z^\beta \,, \qquad \alpha\,,\beta \in \mathbb{R} \,,
\end{equation}
which depends on two constant\footnote{\label{foot:alpha}Technically, the construction remains valid also when $\alpha$ and $\beta$ depend on combinations of phase-space variables that are conserved along classical trajectories (i.e., on {weakly conserved} quantities, such as $p_i$ or $zp_z$). Along each trajectory the lapse coincides with a constant-parameter member of the family \eqref{eq:OurLapse}, so that the conformal Killing equations apply orbit by orbit. Explicitly, taking $\alpha\to a(\phi,p)$ and writing
$\tilde Q:=Q|_{\alpha\to a}$ for any charge $Q$ obeying
$\partial_t Q+\{Q,\alpha z^{\beta}h\}=0$, one finds
$\partial_t\tilde Q+\{\tilde Q,Nh\}
={\{Q,a\}\,z^{\beta}h}
+{\partial_\alpha Q\,\{a,Nh\}}
 \approx 0 $. The first term is proportional to $h$ and the second vanishes
weakly precisely when $a$ is weakly conserved. The same argument holds for $\beta$.} parameters and the coordinate related to the scale factor $z=a^{3/2}$. This functional form cleanly encompasses the most common time choices found in the literature, ranging from cosmic time ($\beta = 0$) to conformal time ($\beta = 2/3$) and harmonic time ($\beta = 2$). In particular, while $\alpha$ is simply an overall constant, keeping $\beta$ arbitrary allows to rigorously test how our symmetry algebra responds to different choices of time, and determine if, and under what conditions, the physical symmetries can be decoupled from time-diffeomorphisms. 

Using the conformally flat form of the supermetric \eqref{eq:confflatX} and the lapse \eqref{eq:OurLapse}, the gauge-fixed metric $\tilde g=g/N$ takes the
strikingly simple form
\begin{equation}
\label{eq:gtilde_kappa}
\tilde g_{\mu\nu}=\frac{c\,l_p^{3}}{\alpha}\,\e^{-\frac{l_p}{2}\,(2-\beta) X^0}\,\eta_{\mu\nu}\,,
\end{equation}
showing that while $\alpha$ is an overall scale, $\beta$ governs the curvature of $\tilde g$. For $n\ge2$\footnote{We will discuss the case $n=1$ separately.} the metric \eqref{eq:gtilde_kappa} is curved whenever
$\beta\neq2$ and flat only at $\beta=2$, where the
lapse exactly cancels the conformal factor of the supermetric. The gauge with $\beta=2$, which we refer to as
\emph{harmonic gauge}, will play a distinguished role below by allowing to obtain physical symmetries in a simple manner, without contributions from gauge transformations (see deparametrisation in
Section~\ref{sec:lift_deparam}).

By the general analysis of Appendix~\ref{app:ED}, the conserved charges
that are linear in the momenta correspond to projectable conformal Killing
vectors of the lifted metric $G$. Solving \eqref{eq:XiG} for a lapse \eqref{eq:OurLapse} with $\beta\neq2$ and constructing the charges (according to \eqref{eq:LiftedQs}) one finds 
\begin{equation}
\label{eq:charges_generic}
\begin{aligned}
\mathcal{H}&=\alpha z^{\beta}h=\frac{\alpha\,z^{\beta-2}}{8 cl_p^3} \left( 4 \sum_{i} p_i^2 - l_p^2\,z^2\,p_z^2\right) \,,\\
\mathcal{D}&=\mathcal{H} t+\frac{zp_z}{\beta-2}\,,\\
\mathcal{K}&=\frac{4c l_p \,z^{2-\beta}}{\alpha(\beta-2)^{2}}-\frac{zp_z}{\beta-2}t-\frac{1}{2}\mathcal{H}t^{2}\,,
\end{aligned}
\end{equation}
together with translations and rotations
\begin{equation}
\label{eq:charges_generic_iso}
\mathcal{P}_i=p_i\,,\qquad \mathcal{J}_{ij}=\chi^{i}p_j-\chi^{j}p_i\,.
\end{equation}
We use a different (calligraphic) notation to distinguish these lifted (gauge-fixed, and generally $t$-dependent) charges from the conformal charges of
Section~\ref{sec:FLRW}; the two are related by deparametrisation, as we will see in Section~\ref{sec:lift_deparam} (note that the time-independent charges \eqref{eq:charges_generic_iso} already coincide with their counterparts in \eqref{eq:charges_physical}). Evaluating the Poisson brackets of the generators listed above, we find that for \textit{any} generic power exponent $\beta \neq 2$, the algebra closes into a direct sum of two completely decoupled structures:
\begin{equation}\label{eq:sliso}
\mathfrak{g}_{\beta \neq 2} \cong \mathfrak{sl}(2, \mathbb{R}) \oplus \mathfrak{iso}(n) \,.
\end{equation}
The $\mathfrak{sl}(2, \mathbb{R})$ subalgebra is spanned by the set $\{\mathcal{H}, \mathcal{D},\mathcal{ K}\}$, reflecting the conformal isometries of the time-coordinate line, where these charges act as generators of M\"obius reparametrisations of the unphysical coordinate time $t$ \cite{sl2r1,sl2r2,sl2r3,EteraDaniele}. On the other hand, the $\mathfrak{iso}(n)$ sector is generated by the matter momenta $\mathcal{P}_i$ and the field-space rotations $\mathcal{J}_{ij}$, forming the standard Euclidean isometry algebra of the $n$-dimensional flat field space. Nothing else survives in the projectable sector for $\beta\neq 2$. Note that the charge $P_z=zp_z$ of \eqref{eq:charges_physical} is not an
independent generator here---it appears only inside $\mathcal{D}$ and $\mathcal{K}$. More
generally, the explicit form of the charges \eqref{eq:charges_generic}
is gauge-dependent (for instance $\mathcal{K}$ carries a term
$\propto z^{2-\beta}$), even though the algebra they close is the same
for every $\beta\neq2$. In fact, one quickly shows that $\mathcal{K}$ is not a Dirac observable (cf.~\eqref{eq:Dirac_observable}), and gauge-invariant physical charges can only be recovered after deparametrisation, as we will see in
Section~\ref{sec:lift_deparam}.

Our general construction, encompassing the family of lapses \eqref{eq:OurLapse}, contains previous implementations of the Eisenhart--Duval lift as special cases, typically performed in a single fixed gauge---most commonly cosmic time, which corresponds here to setting $\alpha = 1$ and $\beta = 0$ \cite{EteraDaniele}. For $n\ge2$, this gauge (like any
$\beta\neq2$ gauge) yields only \eqref{eq:sliso}. We will show that the considerably richer Schr\"odinger symmetry found
with the same gauge in the single-field case in \cite{EteraDaniele} is therefore not a property of cosmic time but of $n=1$, which is geometrically special since \eqref{eq:gtilde_kappa} is flat for \emph{every}
$\beta$.

\subsection{The harmonic gauge and the Schr\"odinger enhancement}
\label{sec:lift_conformal}

The charges \eqref{eq:charges_generic} show a degeneracy at $\beta=2$, where the divergence of some coefficients signals that the harmonic gauge needs to be treated as a distinct case. Indeed, for $\beta=2$, the gauge-fixed metric \eqref{eq:gtilde_kappa} becomes flat, and the projectable algebra expands remarkably. Following again the ED lift prescription of Appendix \ref{app:ED}, one finds the rich family of conserved charges:
\begin{equation}
\begin{aligned}    
\label{eq:charges_harmonic}
\mathcal{H}   &= \alpha z^{2} h\,,\\
\mathcal{D}   &= \mathcal{H}t-\tfrac12\Big(zp_z\ln z+\textstyle\sum_i\chi^{i}p_i\Big)\,,\\
\mathcal{K}   &= -\frac{cl_p^{3}}{4\alpha}\Big(\textstyle\sum_i(\chi^{i})^{2}-\frac{4}{l_p^{2}}\ln^{2}\!z\Big)
         +\tfrac12\Big(zp_z\ln z+\textstyle\sum_i\chi^{i}p_i\Big)t-\tfrac12\mathcal{H}t^{2}\,,\\
\mathcal{P}_0 &= -\tfrac{l_p}{2}zp_z\,,\\
\mathcal{P}_i &= p_i\,,\\
\mathcal{B}_i &= \tfrac{l_p}{2}zp_z\chi^{i}+\tfrac{2}{l_p}p_i\ln z\,,\\
\mathcal{J}_{ij}&= \chi^{i}p_j-\chi^{j}p_i\,,\\
\mathcal{G}_0 &= -\tfrac{l_p}{2}zp_zt-\frac{2cl_p^{2}}{\alpha}\ln z\,,\\
\mathcal{G}_i &= p_it-\frac{cl_p^{3}}{\alpha}\chi^{i}\,.
\end{aligned}
\end{equation}
The first three charges realise the same $\mathfrak{sl}(2,\mathbb{R})$ algebra as in the gauges with $\beta \neq 2$ (generating M\"obius reparametrisations of the
coordinate time) and similarly, the charges $\mathcal{P}_i$ and $\mathcal{J}_{ij}$ generate $\mathfrak{iso}(n)$ as before. However, new charges have appeared, so that the algebra goes beyond the structure \eqref{eq:sliso} of the previous section. The
charges $\mathcal{B}_i$ are time-independent, and are precisely the charges
$B_i$ of Section~\ref{sec:FLRW} (cf.~Eq.~\eqref{eq:charges_physical}); they are boosts mixing the scale factor with the matter fields. Being
Killing vectors of the flat $\tilde g$, they lift trivially
(see Appendix~\ref{app:ED}) and hence survive in the harmonic-gauge lift. The charges $\mathcal{G}_\mu$ are instead new, with no counterpart as isometries of
$\tilde g$: they are a novel $t$-dependent product of the ED lift (the $u$-linear conformal Killing vectors of
Appendix~\ref{app:ED}), analogues of Galilean boosts.  

Evaluating the Poisson brackets, the charges \eqref{eq:charges_harmonic}
close into a centrally extended \emph{Lorentzian}\footnote{The name indicates that the orthogonal factor is the non-compact $\mathfrak{so}(n,1)$---the Lorentz algebra of the $(n+1)$-dimensional minisuperspace---rather than the compact rotations $\mathfrak{so}(n)$ of the ordinary
Schr\"odinger algebra.} Schr\"odinger algebra,
\begin{equation}
\label{eq:sch}
\widehat{\mathfrak{sh}}(n,1)=\big(\mathfrak{sl}(2,\mathbb{R})\oplus\mathfrak{so}(n,1)\big)\ltimes
\big(\mathbb{R}^{\,n+1}_{\mathcal{P}} \oplus \mathbb{R}^{\,n+1}_{\mathcal{G}}\big)
\,,
\end{equation}
where the notation $\widehat{\mathfrak{sh}}(n,1)$ denotes the presence of the central extension which modifies the underlying vector space structure $\mathfrak{sh}(n,1) \oplus \mathbb{R}$ with nontrivial bracket relations. Specifically, the nonvanishing brackets, grouped by sector, are:
\begin{equation}\label{eq:Salg}
\begin{gathered}
\{\mathcal{D},\mathcal{H}\}        =-\mathcal{H}\,,\qquad \{\mathcal{D},\mathcal{K}\}=\mathcal{K}\,,\qquad \{\mathcal{H},\mathcal{K}\}=-\mathcal{D}\,,\\
\{\mathcal{D},\mathcal{P}_\mu\}=-\tfrac12\mathcal{P}_\mu\,,\qquad \{\mathcal{H},\mathcal{G}_\mu\}=\mathcal{P}_\mu\,, \qquad \{\mathcal{D},\mathcal{G}_\mu\}=\tfrac12\mathcal{G}_\mu\,,\qquad \{\mathcal{K},\mathcal{P}_\mu\}=\tfrac12\mathcal{G}_\mu\,,\\
\{\mathcal{P}_\mu,\mathcal{G}_\nu\}=\frac{cl_p^{3}}{\alpha}\,\eta_{\mu\nu}\,,
\end{gathered}
\end{equation}
together with the Lorentz sector $\mathfrak{so}(n,1)$, generated by $\mathcal{M}_{\mu\nu}$---comprising the rotations $\mathcal{M}_{ij}=\mathcal{J}_{ij}$ and boosts $\mathcal{M}_{0i}=\mathcal{B}_i$---which transforms $\mathcal{P}_\mu$ and $\mathcal{G}_\mu$ as vectors, and commutes with
the $\mathfrak{sl}(2,\mathbb{R})$ sector. Just like in Section \ref{sec:FLRW}, $\mu,\nu=0,1,\dots,n$ label the minisuperspace
directions. The last line of \eqref{eq:Salg} is the
\emph{central extension}---the analogue of the mass term in the Bargmann (massive Galilei)
algebra \cite{Bargmann1954}. Its value $cl_p^{3}/\alpha=V_0/\alpha$ is the fiducial comoving
volume $V_0$ measured in Planck units (recall $c=V_0/l_p^{3}$), dressed by
the gauge constant $\alpha$. In the dimensionless normalisation of
Ref.~\cite{EteraDaniele} it is precisely their central charge $2c$ (appearing there in the bracket $\{B_\mp,P_\pm\}$).

That a flat (gauge-fixed) metric should yield a Schr\"odinger algebra is, in
fact, an expected outcome:
the Schr\"odinger group is the maximal group of dynamical symmetries of a
free system~\cite{Niederer:1972zz,Hag,deAlfaro:1976vlx}, realised geometrically
as the conformal transformations of a flat Bargmann space that preserve its
covariantly constant null direction~\cite{Duval:1984cj,Cariglia:2014ysa}. The emphasis here is therefore not on the algebra
\eqref{eq:sch} itself, but on {which} gauges flatten $\tilde g$
 (the harmonic gauge for $n\ge2$; see below for $n=1$).

Crucially, we stress that the charges \eqref{eq:charges_harmonic}---and the algebra
\eqref{eq:sch}---are not gauge-invariant. The charges depend explicitly on the
coordinate time $t$ and on the chosen lapse, and do not all weakly Poisson-commute with the constraint (for example, $\mathcal{K}$ and $\mathcal{G}_\mu$ are not Dirac observables, cf.~\eqref{eq:Dirac_observable}). As a result, they intertwine genuine
symmetries of the system with time reparametrisations (or diffeomorphisms), which are gauge transformations. The same caveat applies to the
one-field Schr\"odinger symmetry of \cite{EteraDaniele}, as discussed below. The physical, gauge-invariant content is recovered only by
deparametrisation, which we carry
out in Section~\ref{sec:lift_deparam}.

\subsubsection*{The one-field case}

For $n=1$ the harmonic gauge loses its distinguished status because the
gauge-fixed metric \eqref{eq:gtilde_kappa} is flat for \emph{every} $\beta$. This is a feature
of two dimensions, where the curvature of a conformally flat metric
$\tilde g_{\mu\nu}=\e^{2\omega}\eta_{\mu\nu}$ is fully encoded in the
Ricci scalar, $R=-2e^{-2\omega}\Box_\eta\omega$. Flatness then reduces to
the single condition that $\omega$ be harmonic, $\Box_\eta\omega=0$. From
\eqref{eq:gtilde_kappa}, $\omega=-\tfrac{l_p}{4}(2-\beta)X^0$ is affine in
the coordinate $X^0$ and therefore harmonic for \emph{any} $\beta$. In
three or more dimensions ($n\ge2$), flatness is strictly stronger than
conformal flatness: the curvature of $\e^{2\omega}\eta$ does not vanish for
a generic harmonic $\omega$, and one finds a
Riemann tensor $\propto(2-\beta)^{2}$, showing that flatness is restored only at
$\beta=2$. This again demonstrates the exceptional nature of two dimensions (here a minisuperspace for an FLRW universe with a single scalar field) already mentioned at the end of
Section~\ref{sec:FLRW}.

Because $\tilde g$ is always flat in two dimensions, the ``Schr\"odinger enhancement'' \eqref{eq:sch}
persists in all gauges (including $\alpha=1$ and $\beta=0$). Indeed, for $n=1$, one recovers the centrally extended
two-dimensional Schr\"odinger algebra
$\mathfrak{sh}(1,1)=\big(\mathfrak{sl}(2,\mathbb{R})\oplus\mathfrak{so}(1,1)\big)\ltimes(\mathbb{R}^2\oplus\mathbb{R}^2)$
of \cite{EteraDaniele} (denoted $\mathfrak{sh}(2)$ there). Its time-independent sector coincides with the isometries already mentioned in Section \ref{sec:FLRW}---the light-cone translations $P_\pm$ and the Lorentz boost $J$ of \eqref{eq:n1NO}. The lift supplements these with the explicitly $t$-dependent charges, namely the $\mathfrak{sl}(2,\mathbb{R})$ generators and the
Galilean boosts $\mathcal{G}_i$ (respectively denoted $Q_{\pm,0}$ and $B_\pm$ in \cite{EteraDaniele}), which enhance the algebra to $\mathfrak{sh}(1,1)$.\footnote{Of course, charges may take different forms in different gauges (although the abstract Schr\"odinger algebra remains the same). For example, in the harmonic gauge ($\beta=2$) the conformal coordinates $X^\mu$
are Cartesian for the flat metric $\tilde g$, so our translations $\mathcal{P}_\mu$
are linear in the momenta. Conversely, in the gauge $\beta=0$ of \cite{EteraDaniele} the Cartesian coordinates are the exponential $x^\pm=z\,\e^{\pm l_p\chi/2}$, so the corresponding translations carry exponential factors.} Note that our general-$n$ construction clarifies why the field momentum $J$ is naturally interpreted as a boost generator: it is the $\mathfrak{so}(1,1)$ survivor of the Lorentz factor $\mathfrak{so}(n,1)$, realised by $p_\chi$ itself in the gauge $\beta=0$ and by $\mathcal{B}_1$ in the harmonic gauge. In the single-field limit the rotation subalgebra $\mathfrak{so}(n)$ disappears, leaving only this boost.

\subsection{Deparametrisation and gauge-invariant content}\label{sec:lift_deparam}

While the ED lift successfully uncovers explicitly $t$-dependent dynamical symmetries, it does not make manifest \emph{all} the symmetries of the free minisuperspace (for example, missing completely the conformal transformations that mix the scale factor with the scalar fields of \eqref{eq:charges_physical}). Indeed, the special conformal generators of $\mathfrak{conf}(n,1)$ lift to non-projectable
conformal Killing vectors, with charges of higher order in the momenta, and so lie outside the projectable set \eqref{eq:charges_harmonic}---in every gauge (see Appendix~\ref{app:ED} for details). Moreover, as stressed in the previous sections, the lifted charges mix physical symmetries with time diffeomorphisms, and are not all Dirac observables. To isolate the physical content, we must disentangle the two by deparametrisation (see Section~\ref{sec:Constrained}): we trade the coordinate time for a physical clock and retain only what is independent of that choice. Crucially, the missing conformal symmetries re-emerge upon deparametrisation.

In a nutshell, {replacing $t$ by the reading of a physical clock turns a gauge-dependent charge into a gauge-invariant Dirac observable}. Concretely, the equations of motion let one solve for the coordinate time as a phase-space function, $t=T(\phi,p)$;
substituting this into a conserved charge $Q(\phi,p,t)$ gives the $t$-independent quantity $\bar Q(\phi,p):=Q\big(\phi,p,T(\phi,p)\big)$, which
weakly commutes with the constraint,
\begin{equation}
\label{eq:dep_dirac}
\{\bar Q, N h\}=\{Q,Nh\}\big|_{t=T}+\frac{\partial Q}{\partial t}\Big|_{t=T}\{T,Nh\}
\;\approx\;\Big(\{Q,Nh\}+\frac{\partial Q}{\partial t}\Big)\Big|_{t=T}\;=\;0\,,
\end{equation}
where we used $\{T,Nh\}=\dot T\approx1$ and the conservation of $Q$. Thus $\bar Q$ is a genuine
relational observable, as expected from the general analysis of Section \ref{sec:Constrained}. In practice, the construction is cleanest when the
lapse is \emph{adapted} to the clock, $\dot T=1$, so that $t=T-T_0$ (where the
integration constant $T_0$ can be set
to zero without loss of generality\footnote{The deparametrised charges for different $T_0$
differ only by terms proportional to the Hamiltonian constraint.}). 

We now work in the harmonic gauge ($\beta=2$) and choose the $n$-th matter field as clock $T=\chi^{n}$, in
parallel with what we did above Eq.~\eqref{eq:FLRW_Hr_Matter}. The corresponding adapted lapse \eqref{eq:OurLapse} is fixed by
\begin{equation}\label{eq:alphachin}
    \dot{\chi}^n=1 \quad \Longleftrightarrow \quad \alpha = \frac{cl_p^3}{p_n}\,,
\end{equation}
following the general structure of \eqref{eq:u_multipl}.\footnote{We note, however, that this adaptation is only a convenience: at $\beta=2$ deparametrisation with \emph{any} $\alpha$ returns the same algebra, a generic $\alpha$ merely dresses the charges with extra (weakly) constant
factors (see footnote \ref{foot:alpha}).} Deparametrisation then consists of using \eqref{eq:alphachin}, replacing $t$ by $\chi^n$, and evaluating the charges \eqref{eq:charges_harmonic} on the constraint surface. The time-independent charges are untouched,
$\mathcal P_\mu\mapsto P_\mu$, $\mathcal B_i\mapsto M_{0i}$,
$\mathcal J_{ij}\mapsto M_{ij}$, and simply get re-read as the conformal
charges of Section~\ref{sub:FLRW_Symmetries}. For the time-dependent ones, on the other hand, deparametrisation explicitly transforms the charges. The sharpest example
is given by the Galilean sector: 
\begin{equation}
\label{eq:worked}
\mathcal G_i \;=\; p_i\,t-\frac{cl_p^3}{\alpha}\,\chi^i
\quad\longmapsto\quad
p_i\,\chi^{n}-p_n\,\chi^{i}\;=\;M^{n}{}_{i}\,,
\end{equation}
showing that the time-dependent Galilean boost of $\chi^i$ has
become a rotation in the $(\chi^i,\chi^n)$-plane of field space. We write the image with the clock index raised (with $\eta^{\mu\nu}$ as in Section~\ref{sub:FLRW_Symmetries}) to keep all signs positive. Moreover, one finds
\begin{equation}
\label{eq:workedG0}
\mathcal G_0 \;=\; -\tfrac{l_p}{2}\,zp_z\,t-\frac{2cl_p^{2}}{\alpha}\ln z
\quad\longmapsto\quad
-\Big(\tfrac{l_p}{2}\,zp_z\,\chi^{n}+\tfrac{2}{l_p}\,p_n\ln z\Big)\;=\;M^{n}{}_{0}\,,
\end{equation}
so that the Galilean boost of the $z$ direction becomes the genuine Lorentz boost mixing the scale factor with the clock field. Writing $\mathcal G_\mu=t\,P_\mu-\tfrac{cl_p^3}{\alpha}X_\mu$, one has $\mathcal G_\mu\;\longmapsto\;X^{n}P_\mu-P^{n}X_\mu\;=\;M^{n}{}_{\mu}$, which extends uniformly to the clock direction itself,
$\mathcal{G}_n=p_n(t-\chi^n)\mapsto 0$,  consistently showing that one cannot boost the clock
relative to itself. Similarly, projecting onto the constraint surface, the special conformal charge $\mathcal{K}$ of \eqref{eq:charges_harmonic} becomes
\begin{equation}
\label{eq:workedK}
\mathcal K\;\longmapsto\;
\tfrac12\,\chi^{n}\big(zp_z\ln z+\chi^ip_i\big)-\tfrac{p_n}{4}\Big(\textstyle\sum_i(\chi^i)^2-\tfrac{4}{l_p^2}\ln^2 z\Big)
\;=\;\tfrac12\,K^{n}\,,
\end{equation}
which is nothing but one of the conformal charges of Section~\ref{sub:FLRW_Symmetries}, specifically the one along the $\chi^n$ clock direction (here $K^{n}=K_{n}$, the clock index being spacelike). The rest of the $\mathfrak{sl}(2,\mathbb R)$ sector is simple: $\mathcal H\mapsto0$ and $\mathcal D\mapsto-\tfrac12D$, which is again the (physical) dilatation of \eqref{eq:charges_physical}. Collecting everything, one has
\begin{equation}
\label{eq:mmap}
\begin{gathered}
\mathcal P_\mu\mapsto P_\mu\,,\qquad
\mathcal B_i\mapsto M_{0i}\,,\qquad
\mathcal J_{ij}\mapsto M_{ij}\,,\qquad
\mathcal G_\mu\mapsto M^{n}{}_{\mu}\,,\\[2pt]
\mathcal H\mapsto0\,,\qquad
\mathcal D\mapsto-\tfrac12\,D\,,\qquad
\mathcal K\mapsto\tfrac12\,K^{n}\,,
\end{gathered}
\end{equation}
where the identities in the first line hold exactly and those in the second hold on $\mathcal{C}_H$. Note that the images of the Galilean sector are redundant: $\mathcal G_0\mapsto M^{n}{}_{0}=-M_{0n}$ duplicates, up to sign, the image of $\mathcal B_n$, and $\mathcal G_i\mapsto M^{n}{}_{i}=-M_{in}$ those of $\mathcal J_{in}$.

Nothing hinges on the matter character of the clock. Indeed, deparametrising
instead with respect to the variable $X^0=-\tfrac2{l_p}\ln z$,
whose adapted lapse has $\alpha=2cl_p^2/(zp_z)$ such that $\dot X^0=1$, the procedure returns the same map \eqref{eq:mmap} with $n$ replaced by $0$: the recovered
special conformal charge is $\mathcal K\mapsto\tfrac12K^{0}=-\tfrac12K_{0}$
(the lowered-index sign reflecting the timelike character of $X^0$), and
the boost roles swap automatically, $\mathcal G_0\mapsto M^{0}{}_{0}=0$
while $\mathcal G_i\mapsto M^{0}{}_{i}=-M_{0i}$.

With either clock, the deparametrised charges comprise the translations
$P_\mu$, the full Lorentz sector $M_{\mu\nu}$, the dilatation $D$, and
exactly one special conformal charge---$K_n$ or $K_z$---the images of the
Galilean sector merely duplicating Lorentz generators already present. The
remaining $n$ special conformal charges cannot be produced this way: as
noted at the beginning of this section, their lifts are non-projectable in
every gauge. Nevertheless, the deparametrised charges are weak Dirac observables and form a closed Poisson algebra. Evaluating their brackets therefore generates the remaining special conformal charges as
\begin{equation}
\label{eq:closureK}
\{K_{n},M_{\mu n}\}=K_\mu\,, \qquad \{K_0,M_{0i}\}=K_i \,,
\end{equation}
for the matter and geometric clock respectively (where $\mu \neq n$, and $K_z=K_0$ in the covariant notation of \eqref{eq:MDKcharges}). The
deparametrised charges therefore \emph{generate} the whole of
$\mathfrak{conf}(n,1)$\footnote{Of the $\dim (\widehat{\mathfrak{sh}}(n,1))=3+{n(n+1)}/2+2(n+1)+1$ lifted generators, two are mapped to zero (the Hamiltonian constraint and the clock Galilean boost), while the central charge is mapped by \eqref{eq:alphachin} to the clock momentum $p_n=P_n$ and is thereby absorbed into the translation sector, ceasing to be an independent (central) generator. The $n$ surviving Galilean images duplicate Lorentz generators and are traded for the $n$ special conformal charges of \eqref{eq:closureK}, so that $\dim(\mathfrak{conf}(n,1))=\dim (\widehat{\mathfrak{sh}}(n,1))-3={(n+2)(n+3)}/2$.} as in \eqref{eq:conf_algebra}. In this precise sense, the lift---used in the
harmonic gauge and deparametrised---recovers the complete conformal
symmetry of Section~\ref{sec:FLRW}, including the
relational-time-dependent charges that the clock-reduced descriptions of
Section~\ref{sub:FLRW_Symmetries} could not see.

Interestingly, the original $\mathfrak{sl}(2,\mathbb R)$ does not survive deparametrisation: while $\mathcal H$ vanishes, $\mathcal D$ and $\mathcal K$ cease to generate M\"obius reparametrisations of the
coordinate time, landing instead on the conformal dilatation and a special
conformal charge. This is precisely what one should expect: the
$\mathfrak{sl}(2,\mathbb R)$ structure encountered in \eqref{eq:sliso} and \eqref{eq:sch} acts on the unphysical time
parameter itself; once that parameter is eliminated, only its induced action on the physical phase space survives.

To complete the picture, we comment on the fact that, for
$\beta\neq2$, the lifted algebra is only
$\mathfrak{sl}(2,\mathbb R)\oplus\mathfrak{iso}(n)$---there are no boosts
$\mathcal B_i$ and no Galilean sector to begin with---so deparametrisation
can return at most a small subalgebra of $\mathfrak{conf}(n,1)$. For instance, taking the scale factor itself as clock,
\begin{equation}
   \dot z=1  \quad \Longleftrightarrow \quad \alpha=-\frac{4cl_p}{zp_z} \,, \quad \beta=1 \,, 
\end{equation}
yields
$\mathcal H\mapsto0$, $\mathcal D\mapsto-P_z$ and $\mathcal K\mapsto0$,
leaving only the generators $\{P_z,\,p_i,\,J_{ij}\}$, namely the Euclidean algebra
$\mathfrak{iso}(n)$ together with the homothety charge---precisely the
geometric-clock reduction of Section~\ref{sub:FLRW_Symmetries} (see the discussion below \eqref{eq:conf_algebra}). Closure
cannot help here: $P_z$ commutes with $p_i$ and $J_{ij}$, so the set is
already closed, and lacking both a special conformal seed and a Lorentz
boost, no sequence of brackets can climb back to $\mathfrak{conf}(n,1)$. The obstruction is therefore not algebraic but geometric: in the curved gauge the missing conformal generators are non-projectable (using the language of Appendix \ref{app:ED}). The full set of charges is recovered only for appropriate clocks and their associated gauge,\footnote{Recall that the ED lift requires a non-degenerate Lagrangian, which the deparametrised theory does not admit (see discussion
below \eqref{eq:Reduced_Action_Velocities}). This is why the construction runs on
the unreduced, gauge-fixed system, with deparametrisation applied afterwards to the charges rather than to the theory.} here adapting to the flat structure of the gauge-fixed geometry. Indeed, $\ln z$ is linear in the Cartesian coordinate $X^0$ of $\tilde g$, whereas $z$ is exponential in it. For the free models considered in this paper such an adapted clock always exists, since the supermetric is conformally flat and the harmonic gauge is available. Whether an analogous construction persists for more general cosmological models remains an open question.

To summarise, while the physical dynamical symmetry of the free FLRW
minisuperspace is the conformal algebra $\mathfrak{conf}(n,1)$, what changes from gauge to gauge is
instead the symmetry algebra made manifest by the Eisenhart--Duval lift. This is reduced in general to $\mathfrak{sl}(2,\mathbb{R})\oplus\mathfrak{iso}(n)$, while in the harmonic gauge it enlarges to the Schr\"odinger algebra.
In all such cases, the gauge-fixed charges are conserved but not Dirac observables in general. Their physical content is recovered only after deparametrisation, which removes the dependence on the coordinate time and maps them onto the relational conformal charges of Section~\ref{sec:FLRW}. The harmonic gauge is therefore distinguished not because it introduces additional physical symmetries, but because, by adapting to the physical clocks available in this context, it enables the deparametrised lifted charges to recover the full physical conformal algebra.

This also clarifies the role of the Eisenhart--Duval lift anticipated at the end of Section~\ref{subsec:Supermetric}. Even when a reduced relational description reveals only the isometries of the reduced metric associated with the chosen clock, the lift systematically reconstructs relational observables by geometrising their explicit time dependence. Indeed, for a system given a priori in the square-root form \eqref{eq:Relational_Hamiltonian_Supermetric}, reinstating the clock as a configuration variable and fixing the clock-adapted lapse \eqref{eq:u_multipl} returns a member of the family \eqref{eq:OurLapse}. For a matter clock this is the harmonic gauge, with the $\alpha$ of \eqref{eq:alphachin}, and the lift then returns the full conformal algebra already obtained directly from the minisuperspace geometry. For the scale factor $z$ it is instead the curved gauge $\beta=1$, and only a subalgebra is recovered, as we saw above. Ultimately, the lift yields a way to construct relational observables, although whether it captures all of them depends on the choice of clock. Unlike the direct conformal analysis, however, it also applies to more general cosmological models with potential terms.

\section{Conclusion and outlook}\label{sec:outlook}

In this paper we have outlined a framework for classifying the dynamical symmetries of time-reparametrisation invariant minisuperspace models in a way that distinguishes gauge from physical content, and applied it in full detail to flat FLRW cosmology minimally coupled to $n$ free massless scalar fields. We constructed the conformal Killing vectors of the supermetric and showed that the associated Noether charges realise, as weak Dirac observables, the maximal conformal algebra $\mathfrak{conf}(n,1)\simeq\mathfrak{so}(n+1,2)$, extending single-field results to an arbitrary number of scalar fields. We then employed the Eisenhart--Duval lift for a family of lapse functions rather than a single fixed gauge, obtaining new coordinate-time-dependent charges. We showed, however, that such an extension is gauge dependent. Generic gauges yield only $\mathfrak{sl}(2,\mathbb{R})\oplus\mathfrak{iso}(n)$, while the harmonic gauge, which flattens the gauge-fixed metric, enhances this to a centrally extended Schr\"odinger algebra $\widehat{\mathfrak{sh}}(n,1)$. Deparametrising the resulting charges we showed that in every gauge they land inside the gauge-invariant conformal algebra obtained directly from the minisuperspace CKV analysis, and that they reproduce it in full only in the harmonic gauge, where even the relationally time-dependent charges, that no clock-reduced description makes manifest, are recovered. The previously reported Schr\"odinger symmetry is therefore obtained as a property of one specific time gauge. The single-field case, in which the gauge-fixed metric is flat for every lapse, is the two-dimensional exception that explains why it was found there in the first place. We also showed (in Appendix~\ref{app:reconstruction}) that the conformal charges form an overcomplete set of Dirac observables, from which the relational
trajectories can be reconstructed algebraically,
without integrating any equation of motion.

Nonetheless, the minisuperspace CKV analysis relies essentially on the vanishing of the potential, so that the dynamics follows null minisuperspace geodesics, which is what allows the CKVs of the supermetric to generate the full set of physical symmetries. Once a potential is present this construction no longer applies, and the ED lift becomes the necessary tool, since its construction holds for an arbitrary potential from the outset (see Appendix~\ref{app:ED}). Such potentials arise generically once one departs from minimally coupled massless fields in a flat, homogeneous and isotropic universe. For instance, a nontrivial potential can be generated directly by a scalar field potential or non-minimal coupling in the matter sector, by spatial curvature, or by the shear degrees of freedom of an anisotropic Bianchi model (which act as an effective potential for the isotropic sector). 

We note that one anisotropic model is in fact automatically covered by this work. A Bianchi~I universe carries no spatial curvature and therefore generates no potential; writing the directional scale factors in the Misner parametrisation $a_1 = z^{2/3} e^{\beta_+ + \sqrt{3}\beta_-}$, $a_2 = z^{2/3} e^{\beta_+ - \sqrt{3}\beta_-}$, $a_3 = z^{2/3} e^{-2\beta_+}$,
so that $z^2 = a_1 a_2 a_3$ is again the spatial volume (see, e.g., \cite{BojoBook}), its two anisotropy degrees of freedom $\beta_\pm$ enter the action exactly as free massless scalar fields, with $\beta_\pm = (l_p/3)\,\chi$ (see also~\cite{Aniso,AC_thesis}). A Bianchi~I universe with $n$ scalar fields is thus equivalent to the model studied here with $n+2$ fields, and its dynamical symmetry algebra is $\mathfrak{conf}(n+2,1)$; the vacuum case yields $\mathfrak{conf}(2,1)\simeq\mathfrak{so}(3,2)$.

We are currently extending the analysis developed here to the remaining Bianchi
types and to spatially curved FLRW geometries via the ED lift (see \cite{Geiller:2022baq} for an analysis on phase space without the ED lift). We stress, however, that the completeness of the classification
obtained in this paper rests on the conformal flatness of the FLRW supermetric,
which is what makes a flattening (harmonic) gauge available in the first place.
This is not guaranteed in more general minisuperspaces, where the lift remains a systematic generator of relational observables but need not, by itself, guarantee that the resulting set is exhaustive. We also note that the argument confining the explicitly time-dependent sector to $\mathfrak{sl}(2,\mathbb{R})$ uses the vanishing of the potential (as described in Appendix~\ref{app:ED}), so that the form this sector takes is no longer fixed a priori once a potential
is switched on.

A second natural target for this framework is black hole mechanics. The symmetries of the black hole interior, and of homogeneous minisuperspace models more generally, have been obtained directly on phase space from the homothetic
Killing vectors of the field-space metric, without recourse to the lift~\cite{Geiller:2020xze,Geiller:2022baq}. Moreover, the ED lift has been used in a fixed time gauge to uncover a Schr\"odinger symmetry for Schwarzschild mechanics in~\cite{EteraDaniele}; the same algebra was also found by means of canonical transformations for spherically symmetric static minisuperspaces in~\cite{Sano:2025xit}. The tools developed here could be used to establish, as we have done for FLRW cosmology, which part of these symmetries is physical and which is a property of the gauge in which they were found. A related separation between residual diffeomorphisms and a generalised Schr\"odinger symmetry has been observed for Vaidya superspace~\cite{Ribisi:2024tmk}.

We stress that the gauge-invariant algebra $\mathfrak{conf}(n,1)$ characterises the physical dynamics itself rather than the variables used to present it, so that any system reproducing it shares that dynamics. This is the perspective behind the proposal of hydrodynamics on (mini)superspace~\cite{Oriti:2024qav,BenAchour:2024gir}, in which the same symmetry structure relates cosmological dynamics to the hydrodynamics of quantum fluids. In this view, the universe is pictured as a quantum-gravity condensate \cite{ORITI2017235,GFTquantumST_Oriti}, with a mean field hydrodynamic regime obeying a conformally invariant non-linear extension of the Wheeler--DeWitt equation on minisuperspace, and whose symmetries can influence the dynamics of collective excitations \cite{Bogo}. Correspondences of this kind have in fact been established directly at the level of the equations of motion, from isotropic cosmologies mapped onto Bose--Einstein condensates~\cite{Lidsey:2003ze,Lidsey:2013osa}, to Bianchi~I backgrounds with a cosmological constant~\cite{DAmbroise:2010njf} (see also \cite{Gumjudpai:2007bx,Phetnora:2008mf} for scalar-field Friedmann dynamics recast in nonlinear Schr\"odinger form). Our results then specify, in a gauge-invariant way, which symmetry such a system---a condensate in the laboratory, or a conformal field theory---would have to realise in order to reproduce the relational dynamics of a cosmological spacetime.

Within the tradition of minisuperspace quantisation \cite{BojoBook}, our results provide a necessary preliminary to any use of dynamical symmetries to inform the quantum theory. Factor ordering of the Hamiltonian constraint, the choice of physical inner product, or the construction of a preferred Hilbert space representation can all be guided by the symmetry algebra of the classical system (see e.g., \cite{Rostami:2015kua}); but such guidance is only meaningful if the algebra in question is genuinely physical rather than an artefact of a gauge choice. Our results show that the correct input for these quantisation choices is the conformal algebra, realised as weak Dirac observables independently of any time gauge, rather than the Schr\"odinger algebra found in a fixed gauge. In particular, the fact that the conformal charges are only \emph{weak} Dirac
observables (cf.~\eqref{eq:CKV_conservation}), means that at the quantum level they preserve the physical state space only for
an ordering in which the constraint operator transforms covariantly under conformal rescalings of the supermetric. This singles out the conformal Laplacian on minisuperspace as the natural ordering of the Wheeler--DeWitt operator~\cite{DeWitt:1967yk,Rostami:2015kua}. The conserved quantities of higher order in the momenta, associated with the non-projectable conformal Killing vectors of the lift, would in turn correspond to higher-order operators, and it would be interesting to see what constraints they impose in their own right.

More broadly, it remains an open question whether any part of this analysis can be extended beyond finite-dimensional truncations of superspace. The well-known difficulties discussed in the introduction obstruct a direct extension of our methods to the fully nonlinear and inhomogeneous theory. A more promising direction may be a perturbative or linearised treatment of inhomogeneous fluctuations around a minisuperspace background, along the lines of relational formulations of cosmological perturbation theory \cite{Giesel:2007wi,Giesel:2007wk}, where some of these obstructions may become tractable within a controlled expansion.

We close on the pattern that has recurred throughout. Singling out a clock hides genuine charges, while fixing a lapse displays charges that are not gauge-invariant. The lift makes the first visible again, and deparametrisation disentangles the second, leaving unchanged the true physical algebra, ultimately defined by Dirac observables. A symmetry algebra is, in this sense, only as physical as the observables that carry it.

\section*{Acknowledgements}
The authors are grateful to Etera Livine for fruitful discussions. The authors acknowledge support from Grant PR28/23 ATR2023-145735 funded by MCIN/AEI/10.13039/501100011033
and from the WOST (WithOut SpaceTime) project, supported by Grant ID 63683 from the John Templeton Foundation.
FG acknowledges support from Istituto
Nazionale di Fisica Nucleare (INFN) through the Theoretical Astroparticle Physics (TAsP) project and from Dipartimento di Fisica e Astronomia (DFA) of the University of Padua.

\appendix

\section{Brief review of the Eisenhart--Duval lift}
\label{app:ED}

The Eisenhart--Duval (ED) lift embeds the dynamics of a mechanical system into the null geodesics of a Lorentzian metric living in two additional dimensions~\cite{ED,ED2,ED3}. This has become a standard tool for uncovering hidden symmetries, as it converts questions about dynamical symmetries of a mechanical system into questions about the conformal geometry of the lifted metric \cite{Cariglia:2014ysa}. Its appeal in the present context is clear: the symmetry analysis of
Sections~\ref{sec:Constrained} and \ref{sec:FLRW} relied entirely on classical trajectories being null geodesics of the minisuperspace metric. This structure is, however, special to free time-reparametrisation invariant models: it is lost as soon as a potential is present, and it is obscured once a specific time gauge is adopted. The ED lift restores a null-geodesic formulation in an enlarged space, for any potential and any choice of gauge.

Our starting point is the reparametrisation-invariant minisuperspace action \eqref{eq:Supermetric_action}, where no specific functional form for the lapse $N$ is required at this stage. The entire construction of the ED lift holds for an arbitrary (nonvanishing) lapse function, and a concrete form of $N$ will only be needed to solve the conformal Killing equations explicitly (see below and Section~\ref{sec:Lift}). We emphasise this point since the lift is usually implemented only after committing to a specific gauge from the outset (as e.g., in \cite{EteraDaniele}); keeping $N$ arbitrary is a novel construction that enables the gauge-family analysis of Section~\ref{sec:Lift}. Concretely, we work with the action
\begin{equation}
\label{eq:gf_system}
S_{N}[\phi^a] = \int \d t \left[
\frac{1}{2}\, \tilde g_{ab}(\phi)\,\dot\phi^a \dot\phi^b - \tilde V(\phi)
\right]\,,
\end{equation}
where
\begin{equation}
    \tilde g_{ab}\coloneqq \frac{g_{ab}}{N}\,,
\qquad
\tilde V\coloneqq N V\,.
\end{equation}

The ED lift of \eqref{eq:gf_system} introduces two additional coordinates
$(u,w)$ and the extended metric
\begin{equation}\label{eq:lifted_supermetric}
\d s^2_{\text{ED}}
= G_{AB}\,\d q^A \d q^B
= \tilde g_{ab}(\phi)\, \d\phi^a \d\phi^b
+ 2\, \d u\,\d w
- 2\,\tilde V(\phi)\, \d u^2 ,
\qquad
q^A = (\phi^a,u,w)\,,
\end{equation}
together with the time-reparametrisation invariant action
\begin{equation}
S_{\text{ED}} = \int \d\lambda\,
\frac{1}{2\mathcal{N}}\, G_{AB}(q)\,\dot q^A \dot q^B\,,
\end{equation}
where $\lambda$ is an arbitrary parameter and $\mathcal{N}$ the associated
lapse. The lifted system is free: the potential has been absorbed into the
geometry. Variation with respect to $\mathcal{N}$ imposes a null
constraint, so the lifted dynamics consists of null geodesics of
$G_{AB}$~\cite{EteraDaniele}. The momenta read
\begin{equation}
    p_a = \frac{\tilde g_{ab}\,\dot \phi^b}{\mathcal{N}}\,,\qquad
    p_u = \frac{\dot w- 2\tilde V \dot u }{\mathcal{N}}\,,\qquad
    p_w = \frac{\dot u}{\mathcal{N}}\,,
\end{equation}
and the total Hamiltonian is
\begin{equation}
\label{eq:HED}
    H_{{\rm{ED}}}={\mathcal{N}} \left[ p_u\,p_w
    + \frac{1}{2}\,\tilde g^{ab} \,p_a\,p_b + \tilde V(\phi)\, p_w^2 \right]
    \eqqcolon {\mathcal{N}}\, h_{\rm{ED}}\,.
\end{equation}
Since $u$ and $w$ are cyclic, $p_u$ and $p_w$ are constants of motion, and
we may set $p_w=1$ without loss of generality. The lifted constraint
$h_{\rm ED}=0$ then determines
\begin{equation}
\label{eq:pu_rule}
p_u = -\left(\tfrac{1}{2}\tilde g^{ab}p_ap_b + \tilde V\right)
    = -N\left(\tfrac{1}{2} g^{ab}p_ap_b + V\right) = -N h\,.
\end{equation}
Note that this does \emph{not} impose the constraint $h=0$ of the original
theory: the minisuperspace model is recovered by
restricting to $h= 0$. Deparametrising the lifted equations of motion
with respect to $u$ (using $\d u = \mathcal{N}\d\lambda$) yields
\begin{equation}
    \frac{\d \phi^a}{\d u} = \{\phi^a,N h\}\,,\qquad
    \frac{\d p_a}{\d u} = \{p_a,Nh\}\,,
\end{equation}
which are exactly Hamilton's equations of the system
\eqref{eq:gf_system}. The ED construction is therefore dynamically
equivalent to the original model via the projection
\begin{equation}
\label{projection rules}
    p_u\rightarrow - N\,h\,,\qquad p_w \rightarrow 1\,,\qquad u \rightarrow t\,.
\end{equation}
Let us also note that if the lapse is adapted to an internal clock $T$ as
in Eq.~\eqref{eq:u_multipl}, then $u$ ticks synchronously with $T$ along
the projected trajectories; this observation is exploited in
Section~\ref{sec:Lift} to eliminate the gauge time from the charges obtained from the ED lift, as explained below.

The advantage of the lift is that dynamical symmetries can again be
studied geometrically, now at the level of the lifted metric. A conformal
Killing vector (CKV) of $G_{AB}$,
\begin{equation}
\label{eq:liftedCKE}
\mathcal{L}_\Xi\, G_{AB} = \Omega\, G_{AB}\,,
\qquad
\Xi = \xi^a\partial_a + \Xi^u\partial_u + \Xi^w\partial_w\,,
\end{equation}
generates a conserved charge of the lifted system,
$Q^{\rm ED}_\Xi=\Xi^Ap_A$, by the same computation as in
Section~\ref{sub:FLRW_Symmetries}. Through the projection rules
\eqref{projection rules}, such charges reduce to (generally
$t$-dependent) conserved quantities of the original minisuperspace system,
\begin{equation}\label{eq:LiftedQs}
Q_\Xi = \Big(\Xi^w - N h\,\Xi^u + \xi^a p_a\Big) \Big\vert_{u=t}\,,\end{equation}
satisfying
\begin{equation}
\frac{\d Q_\Xi}{\d t}= \{ Q_\Xi, Nh\}+ \frac{\partial Q_\Xi}{\partial t}=0\,.
\end{equation}

In writing Eq.~\eqref{eq:LiftedQs} we have implicitly assumed that the
components of $\Xi$ carry no dependence on $w$; this assumption selects a preferred class of CKVs, which
we now characterise. The lifted metric admits the covariantly constant
null vector $\partial_w$ (immediate from \eqref{eq:lifted_supermetric},
since $G_{AB}$ is $w$-independent and $G_{Aw}$ is constant), making
\eqref{eq:lifted_supermetric} a metric of Bargmann (pp-wave) type \cite{Duval:1984cj}. CKVs
then split into those preserving this structure (namely with $[\Xi,\partial_w]=0$ so that all components are $w$-independent) which we call \emph{projectable}, and the rest. Only projectable CKVs yield charges of the form \eqref{eq:LiftedQs}; the others are instead associated with conserved quantities of higher order in the momenta~\cite{EteraDaniele}. This restriction reproduces precisely the classical characterisation, given in Refs.~\cite{Duval:1984cj,Duval:1990hj,Duval:2009vt} (see also \cite{Duval:2024eod} for a recent historical review), of the symmetries of a mechanical system as the conformal transformations of its Bargmann lift that
preserve the null direction. We henceforth work in the projectable sector, which captures all conserved charges linear in the momenta.

For projectable CKVs, expanding Eq.~\eqref{eq:liftedCKE} in components yields a few simple structural facts, which classify the projectable sector completely and shape the analysis of
Section~\ref{sec:Lift}.

First, the components along $w$ force
\begin{equation}
\label{eq:rigidity}
\Xi^u = \Xi^u(u)\,,\qquad
\Omega = \partial_u \Xi^u(u)\,,
\end{equation}
which in particular implies that the conformal factor of a projectable CKV \textit{can only depend on
$u$}, in any model and gauge. Moreover, for vanishing potential (the case of interest for this paper), the remaining components bound the possible $u$-dependence polynomially. Specifically, one finds
$\partial_u^2\,\Omega=0$, so that
$\Xi^u\in \text{span }\{1,u,u^2\}$. Since the vector fields $\partial_u$,
$u\partial_u$ and $u^2\partial_u$ generate the M\"obius (projective)
transformations of the time line and close into $\mathfrak{sl}(2,\mathbb{R})$ \cite{sl2r1,sl2r2,sl2r3}, the explicitly $t$-dependent sector of the
lifted symmetries is never larger than $\mathfrak{sl}(2,\mathbb{R})$---in any gauge.

Second, the transverse components of \eqref{eq:liftedCKE} (its purely
minisuperspace $ab$-block) reduce to
\begin{equation}
\label{eq:transverseCKE}
\mathcal{L}_{\xi}\, \tilde g_{ab} = \Omega(u)\;\tilde g_{ab}\,,
\end{equation}
where $\xi\coloneqq\xi^a\partial_a$ is the transverse part of $\Xi$ (possibly
depending parametrically on $u$). This dictates that only vector
fields associated with conformal transformations of
$\tilde g$ with a position-independent factor can appear. This condition is best
read as a statement about the CKVs of the original supermetric. Indeed,
since $\tilde g = g/N$ is conformally related to $g$, the two metrics
possess the same conformal Killing vectors; what changes is the conformal
factor they carry,
\begin{equation}
\label{eq:filter}
\mathcal{L}_\xi\, g = \varphi\, g
\qquad\Longrightarrow\qquad
\mathcal{L}_\xi\, \tilde g = \big[\varphi - \xi^a\partial_a (\ln N)\big]\,\tilde g\,.
\end{equation}
The lapse therefore acts as a \emph{filter} on the conformal algebra of
the minisuperspace: a CKV $\xi$ of $g$ survives the lift if and only if
the gauge-fixed factor $\varphi - \xi^a\partial_a(\ln N)$ is constant over
configuration space, as explained above. When this factor vanishes, $\xi$ is an isometry of
$\tilde g$ and lifts trivially: every Killing vector of $\tilde g$ extends naturally to a Killing vector of the lifted metric, $\Xi=(\xi^a,0,0)$. In other words: isometries always
survive the lift. By contrast, when the
factor is a nonvanishing constant, it must be given by $\Omega(u)$; the vector then
survives solely in combination with a nontrivial transformation of the time coordinate, and its charge carries explicit
$t$-dependence.\footnote{Both situations occur in the model studied in the main text: the homothety $z\partial_z$ of the FLRW supermetric survives in every gauge of the family \eqref{eq:OurLapse}, while the CKVs mixing the scale factor with the matter fields survive precisely in the harmonic gauge, where the lapse exactly cancels their position-dependent factor (see Section~\ref{sec:Lift}).} Thus, different gauges keep different portions of the same conformal algebra, as we discuss explicitly in Section~\ref{sec:Lift}.

Third, the mixed components of \eqref{eq:liftedCKE} close the classification by determining $\Xi^w$ and restricting the transverse vectors $\xi^a$ to be at most linear in $u$. Such a linear part is the only source of transverse charges carrying explicit time dependence, and it exists only for flat gauge-fixed geometries. The reason is geometric: a transverse charge that depends explicitly on time forces the conformal Killing vector to acquire a component along the extra null direction, growing as the time coordinate advances. Consistency then requires that this growth be encoded in a single function on minisuperspace, accounting at once for every transverse direction. Such a function exists only when the transverse direction is one along which the gauge-fixed geometry is strictly invariant---a translation---and, among conformally flat metrics, only the flat ones admit these. Accordingly, on a curved gauge-fixed geometry the projectable charges carrying explicit time dependence are confined to the $\mathfrak{sl}(2,\mathbb{R})$ sector, whereas a flat one supplements them with genuinely transverse time-dependent charges. This is precisely what happens in the harmonic gauge ($\beta=2$ in \eqref{eq:OurLapse}) of the FLRW models,
where the gauge-fixed metric becomes flat and such charges appear, as worked out in Section~\ref{sec:Lift} (see in particular the charges $\mathcal{G}_i$ in \eqref{eq:charges_harmonic}).

\section{Counting charges and reconstructing relational dynamics}\label{app:reconstruction}

The conformal symmetry uncovered in Section~\ref{sec:FLRW} comes with
$(n+2)(n+3)/2$ conserved charges. This is far more than the number of
degrees of freedom of an FLRW universe filled with $n$ scalar fields. We recall that the kinematical phase space, spanned by $(z,\chi^i)$
and their conjugate momenta, is $2(n+1)$-dimensional. The physical phase
space is two dimensions smaller: the theory is fully constrained, so
imposing $h\approx0$ restricts the dynamics to the $(2n+1)$-dimensional
surface $\mathcal C_H$, and identifying the points along each gauge
orbit removes one more dimension~\cite{Gaugebook}. The resulting
$2n$ dimensions are indeed those of the reduced phase space obtained by
deparametrisation in Section~\ref{subs:FLRW_dynamics}. As a result, the $(n+2)(n+3)/2$ charges cannot all be independent, and must
obey functional relations.

Collecting the charges $Q_I$ of Section~\ref{sub:FLRW_Symmetries} and the phase-space coordinates
$Y=(z,\chi^i,p_z,p_i)$ into the Jacobian matrix
\begin{equation}
\label{eq:appB_jacobian}
\mathbb{J}_{IJ}=\frac{\partial Q_I}{\partial Y^J}\,,
\end{equation}
the number of independent charges is counted by its rank. A direct computation gives
\begin{equation}
\label{eq:appB_ranks}
\operatorname{rank}\mathbb{J}=2(n+1)\,,
\qquad
\operatorname{rank}\mathbb{J}\big|_{\mathcal{C}_H}=2n\,.
\end{equation}
Off the constraint surface the rank is maximal: there are as many
independent charges as phase-space coordinates, so the charges
collectively carry the complete kinematical information. On
$\mathcal{C}_H$, two independent values are lost,\footnote{The
restriction is performed \emph{before} differentiating: one solves the
constraint \eqref{eq:ConstrainSurfaceFLRW} for one momentum inside the
charges, and then differentiates with respect to the remaining $2n+1$
coordinates.}  one for each of the two reductions above. In particular, the constraint becomes a
relation among the charges themselves (cf.~\eqref{eq:ConstrainSurfaceFLRW})
\begin{equation}\label{eq:HAMHAM}
    \left(\frac{l_p P_z}{2}\right)^{2} = \sum_{i=1}^n P_i^2 \,,
\end{equation}
and, being constants of motion, the
charges cannot distinguish points along a gauge orbit. They only see the $2n$-dimensional space of orbits, namely the physical phase space. That exactly $2n$ independent constants of motion survive is the hallmark of a fully constrained system: fixing all of them selects a single point of the physical phase
space, and nothing is left to evolve---the dynamics with respect to the time coordinate $t$ is frozen \cite{Kuchar:1991qf,Isham:1992ms}. The physical information is stored instead in the relations that
the fixed values impose among the phase-space variables and, as we now
show, these relations are precisely the relational trajectories of
Section~\ref{subs:FLRW_dynamics}.

Not every choice of $2n$ charges is functionally independent on
$\mathcal C_H$: for instance, no set containing $P_z$ together with all the field momenta can be, precisely because of \eqref{eq:HAMHAM}. As our
$2n$ independent charges we take the momenta of the fields, the $n-1$ rotations sharing the last index (any other set of $n-1$ independent
rotations would serve equally well), and the dilatation:
\begin{equation}
\label{eq:appB_basis}
\big\{P_1,\dots,P_n; J_{1n},\dots,J_{n-1\,n}; D\big\}\,.
\end{equation}
All remaining charges are then determined explicitly: on $\mathcal C_H$, one has \eqref{eq:HAMHAM} as well as
\begin{equation}
\label{eq:appB_relations}
\begin{aligned}
J_{ij}&=\frac{P_j\,J_{in}-P_i\,J_{jn}}{P_n}\,,
\qquad &
B_i &= \frac{2}{l_p P_z}\Big(P_i\, D+\sum_{j=1}^n J_{ij}\,P_j\Big)\,,\\[4pt]
K_z &= \frac{1}{l_p P_z}\Big(D^2+\sum_{i<j} J_{ij}^2\Big)\,,
\qquad &
K_i &= \frac{2}{l_p P_z}\left(B_i D - P_i K_z\right)\,.
\end{aligned}
\end{equation}
While the first and the fourth are purely algebraic manipulations, and valid everywhere on phase space as exact identities, the second and the third hold on $\mathcal C_H$ only: these can be obtained using \eqref{eq:HAMHAM} and Lagrange's identity
$\sum_{i<j}J_{ij}^2=\sum_i(\chi^i)^2\sum_jp_j^2-(\sum_a\chi^ap_a)^2$, respectively.

The equations \eqref{eq:appB_relations} provide the following number of relations: ${(n-1)(n-2)}/{2}$ for the rotations not in
\eqref{eq:appB_basis}, $n$ for the boosts, and $1+n$ for the special
conformal charges. Together with the extra relation \eqref{eq:HAMHAM} one finds 
\begin{equation}
\frac{(n+2)(n+3)}{2}-\underbrace{\left[\frac{(n-1)(n-2)}{2}+n+1+n+1\right]}_{\text{relations}}
=2n
\end{equation}
independent values, in agreement with \eqref{eq:appB_ranks}.

We now turn to the dynamics. Once its value is fixed, each charge becomes a condition on the phase-space variables. Then, inside the $(2n+1)$-dimensional $\mathcal{C}_H$, the $2n$ independent conditions cut out exactly a one-dimensional set, which is the gauge orbit (namely the single point of the physical phase space selected above, now seen
as a curve in $\mathcal C_H$). To read this curve as evolution we must choose a clock, and given the choice in \eqref{eq:appB_basis} it is natural to take $\chi^n$, the field entering every $J_{in}$. Recalling that $J_{in}=\chi^i p_n-\chi^n p_i$ and $D=z p_z \ln z+ \sum_{a=1}^{n}\chi^a p_a $, and making use of the constraint \eqref{eq:HAMHAM}, one arrives at
\begin{equation}
\label{eq:appB_intermediate}
\begin{aligned}
\chi^i(\chi^n)&=\frac{P_i}{P_n}\,\chi^n+\frac{J_{in}}{P_n}\,,
\\
\ln z(\chi^n)&=\frac{1}{P_z}\Big(D-\sum_{i<n}\frac{J_{in}P_i}{P_n}\Big)
-\frac{l_p^2 P_z}{4 P_n}\,\chi^n\,,
\\
p_z(\chi^n)&=\frac{P_z}{z(\chi^n)}\,.
\end{aligned}
\end{equation}
These are still nothing but the definitions of the charges, rearranged;
yet the shape of the relational trajectories of
Section~\ref{subs:FLRW_dynamics} is already recognisable. To complete the identification it is enough to exponentiate the second equation and evaluate all
variables at an arbitrary reference reading $\chi^n_0$ of the clock (which only fixes
where the description is anchored, not the trajectory itself). Then \eqref{eq:appB_intermediate} reproduces the matter-clock trajectories
\eqref{eq:RelationalObsFLRW_Matter} exactly, with
\begin{equation}
\label{eq:appB_constants}
\chi^i_0=\frac{J_{in}+P_i\,\chi^n_0}{P_n}\,,
\qquad
\ln z_0=\frac{1}{P_z}\Big(D-\sum_{i<n} P_i\,\chi^i_0-P_n\,\chi^n_0\Big)\,,
\qquad
p_{z,0}=\frac{P_z}{z_0}\,,
\qquad
p_{i,0}=P_i\,,
\end{equation}
and with the clock momentum $P_n\approx-H_r^{(\text{matter})}$ (cf.~\eqref{eq:deparam_constraint}).

Recall that the quantities $z_0$, $p_{z,0}$,
$\chi^i_0$, $p_{i,0}$ in \eqref{eq:RelationalObsFLRW_Matter} are the $2n$ free integration constants labelling
the general solution, and \eqref{eq:appB_constants}
simply expresses them through the $2n$ charge values. The map is
invertible: evaluating each charge at the reference point gives
$P_i=p_{i,0}$, $P_n\approx-H_r^{(\text{matter})}$,
$J_{in}=\chi^i_0P_n-\chi^n_0P_i$,
$D=z_0p_{z,0}\ln z_0+\sum_a\chi^a_0p_{a,0}$, so the initial
conditions are literally provided by the charges here.

We close with a remark on the choice \eqref{eq:appB_basis}. While that is a
convenient set, it is not unique: any $2n$ charges that remain
functionally independent on $\mathcal C_H$ serve equally well, though
not every set does (for instance, trading one of them for $P_z$ brings
nothing new, since \eqref{eq:HAMHAM} ties $P_z$ to
the momenta). Different admissible choices are naturally adapted to
different clocks. The set $\{P_i, B_i\}$, for example, is equally valid, and here the
inversion is immediate. The definition
$B_i=\tfrac{l_p}{2}P_z\,\chi^i+\tfrac{2}{l_p}\,P_i\ln z$, read as an equation for $\chi^i$ at fixed charge values, is
already the trajectory
\begin{equation}
\label{eq:appB_geom}
\chi^i(z)=\chi^i_0-\frac{4 P_i}{l_p^2 P_z}\,\ln\!\left(\frac{z}{z_0}\right),
\qquad
\chi^i_0=\frac{2B_i}{l_p P_z}-\frac{4 P_i}{l_p^2 P_z}\ln z_0\,,
\end{equation}
which are the geometric-clock trajectories
\eqref{eq:RelationalObsFLRW_Geom}. No new information is contained here:
these are the same physical curves, written as functions of $z$ rather
than of $\chi^n$. The choice of independent charges then mirrors the choice of clock: either basis, like either clock, encodes the same gauge-invariant content.

\bibliographystyle{jhep}
\bibliography{references.bib}

\end{document}